\documentclass[fleqn,usenatbib]{mnras}

\usepackage{mathptmx}
\usepackage{amsmath}
\usepackage{txfonts}

\usepackage[T1]{fontenc}

\usepackage{graphicx}
\usepackage{xspace}
\usepackage{tabu}
\usepackage{makecell}
\usepackage{txfonts}
\DeclareRobustCommand{\VAN}[3]{#2}
\let\VANthebibliography\thebibliography
\def\thebibliography{\DeclareRobustCommand{\VAN}[3]{##3}\VANthebibliography}

\usepackage{graphicx}	% Including figure files
\usepackage{amsmath}	% Advanced maths commands

\newcommand{\gomez}{G\'{o}mez}

\newcommand{\todo}[1]{\iffalse #1 \fi}

\newcommand{\mj}{$M_{\mathrm{Jup}}$\xspace}

\newcommand{\gaia}{\emph{Gaia}\xspace}
\newcommand{\hip}{\emph{Hipparcos}\xspace}
\newcommand{\vlmsc}{\texttt{very\_low\_mass\_stellar\_companion}\xspace}
\newcommand{\bs}{\texttt{binary\_star}\xspace}
\newcommand{\bdc}{\texttt{brown\_dwarf\_companion}\xspace}
\newcommand{\fpo}{\texttt{false\_positive\_orbit}\xspace}
\newcommand{\exoplanet}{\texttt{exoplanet}\xspace}
\newcommand{\sscc}{preselected sources\xspace}
\newcommand{\bsscc}{high-probability preselected sources\xspace}

\newcommand{\typeOrb}{`Orbital'\xspace}
\newcommand{\typeASB}{`AstroSpectroSB1'\xspace}
\newcommand{\typeSB}{`SB1'\xspace}

\newcommand{\typeOrbAlt}{`OrbitalAlternative'\xspace}
\newcommand{\typeOrbAltVal}{`OrbitalAlternativeValidated'\xspace}

\newcommand{\typeOrbAltStar}{`OrbitalAlternative*'\xspace}
\newcommand{\typeOrbTar}{`OrbitalTargetedSearch'\xspace}
\newcommand{\typeOrbTarVal}{`OrbitalTargetedSearchValidated'\xspace}

\newcommand{\typeOrbTarStar}{`OrbitalTargetedSearch*'\xspace}

\newcommand{\nssSolutionType}{\texttt{nss\_solution\_type}\xspace}
\newcommand{\nssTBO}{\texttt{nss\_two\_body\_orbit}\xspace}
\newcommand{\nssBM}{\texttt{binary\_masses}\xspace}
\newcommand{\gstable}{\texttt{gaia\_source}\xspace}

\title[Machine learning-based identification of exoplanet orbits]{Machine learning-based identification of \gaia astrometric exoplanet orbits}

\author[Sahlmann \& \gomez]{
Johannes Sahlmann$^{1,2}$\thanks{E-mail: Johannes.Sahlmann@esa.int (JS)}
and Pablo \gomez$^{1}$\thanks{E-mail: Pablo.Gomez@esa.int (PG)}
\\
$^{1}$European Space Agency (ESA), European Space Astronomy Centre (ESAC), Camino Bajo del Castillo s/n, 28692 Villanueva de la Ca\~nada, Madrid, Spain\\
$^{2}$RHEA Group for the European Space Agency (ESA), European Space Astronomy Centre (ESAC), Camino Bajo del Castillo s/n,\\ 28692 Villanueva de la Ca\~nada, Madrid, Spain
}

\date{Accepted 2024 December 16. Received 2024 December 13; in original form 2024 April 14}

\pubyear{2024}

\begin{document}
\label{firstpage}
\pagerange{\pageref{firstpage}--\pageref{lastpage}}
\maketitle

\begin{abstract}
The third \gaia data release (DR3) contains $\sim$170\,000 astrometric orbit solutions of two-body systems located within $\sim$500 pc of the Sun. Determining component masses in these systems, in particular of stars hosting exoplanets, usually hinges on incorporating complementary observations in addition to the astrometry, e.g.\ spectroscopy and radial velocities. 
Several \gaia DR3 two-body systems with exoplanet, brown-dwarf, stellar, and black-hole components have been confirmed in this way.   
We developed an alternative machine learning approach that uses only the \gaia DR3 orbital solutions with the aim of identifying the best candidates for exoplanets and brown-dwarf companions. Based on confirmed substellar companions in the literature, we use semi-supervised anomaly detection methods in combination with extreme gradient boosting and random forest classifiers to determine likely low-mass outliers in the population of non-single sources.
We employ and study feature importance to investigate the method's plausibility and produced a list of 20 best candidates of which two are exoplanet candidates and another five are either very-massive brown dwarfs or very-low mass stars. Three candidates, including one initial exoplanet candidate, correspond to false-positive solutions where longer-period binary star motion was fitted with a biased shorter-period orbit. We highlight nine candidates with brown-dwarf companions for preferential follow-up. {The} companion around the Sun-like star G\,15-6 could be confirmed as a genuine brown dwarf using external radial-velocity data. This new approach is a powerful complement to the traditional identification methods for substellar companions among \gaia astrometric orbits. It is particularly relevant in the context of \gaia DR4 and its expected exoplanet discovery yield.

% This is a simple template for authors to write new MNRAS papers.
% The abstract should briefly describe the aims, methods, and main results of the paper.
% It should be a single paragraph not more than 250 words (200 words for Letters).
% No references should appear in the abstract.
\end{abstract}


%%%%%%%%%%%%%%%%%%%%%%%%%%%%%%%%%%%%%%%%%%%%%%%%%%

%%%%%%%%%%%%%%%%% BODY OF PAPER %%%%%%%%%%%%%%%%%%

\section{Introduction}
The third \gaia data release \citep[DR3,][]{GaiaCollaboration:2016aa,2023A&A...674A...1G} delivered the first uniform and large-scale census of astrometric binaries and increased the number of available orbital solutions by two orders of magnitude from a few thousand \citep{Mason:2001fk} to $\sim$170\,000. The revolutionary accuracy of \gaia made it possible to determine a large number of astrometric orbits with semi-major axes smaller than 1 milli-arcsecond (mas) for the first time \citep{2023A&A...674A...9H}, finally opening the astrometric exoplanet-science window widely. However, small semi-major axes do not necessarily imply low companion masses: because the \gaia DR3 astrometric orbits usually refer to the unresolved photocentre of a binary, the principal astrophysical false-positive scenario for an exoplanet candidate is a nearly equal-mass and equal-brightness binary star \citep[e.g.][]{2023AA...674A..34G, 2023AJ....165..266M}.

The astrometric orbits of several known giant exoplanets have been determined in \gaia DR3  \citep{2023AA...674A..10H,2023AA...674A..34G,2022AJ....164..196W}, leading to better constraints on their orbital configurations and masses. \gaia DR3 also contains a number of new giant exoplanet candidates from astrometry.\footnote{The \gaia DR3 exoplanet candidate list is at \url{https://www.cosmos.esa.int/web/gaia/exoplanets}} The default procedure to confirm those is to gather additional data, e.g.\ spectroscopy and precision radial-velocities (RV), that exclude any false-positive scenario.

Here we propose an intermediate step to identify the most promising exoplanet candidates that can be prioritised for follow-up. Our machine learning approach relies only on \gaia DR3 without the need for external datasets. Because of the very small number of confirmed exoplanets with \gaia DR3 orbits, which we need to `train` our models, we use as proof-of-concept the identification of substellar companions in general, i.e.\ both extrasolar planets and brown-dwarf companions. Since the mass distribution of substellar companions is structured but continuous \citep[e.g.][]{Grether:2006kx,Sahlmann:2011fk}, this does not imply a loss of applicability, but it conveniently doubles the size of the `training` sample. 
\begin{figure}
% \centering
\includegraphics[width=\columnwidth]{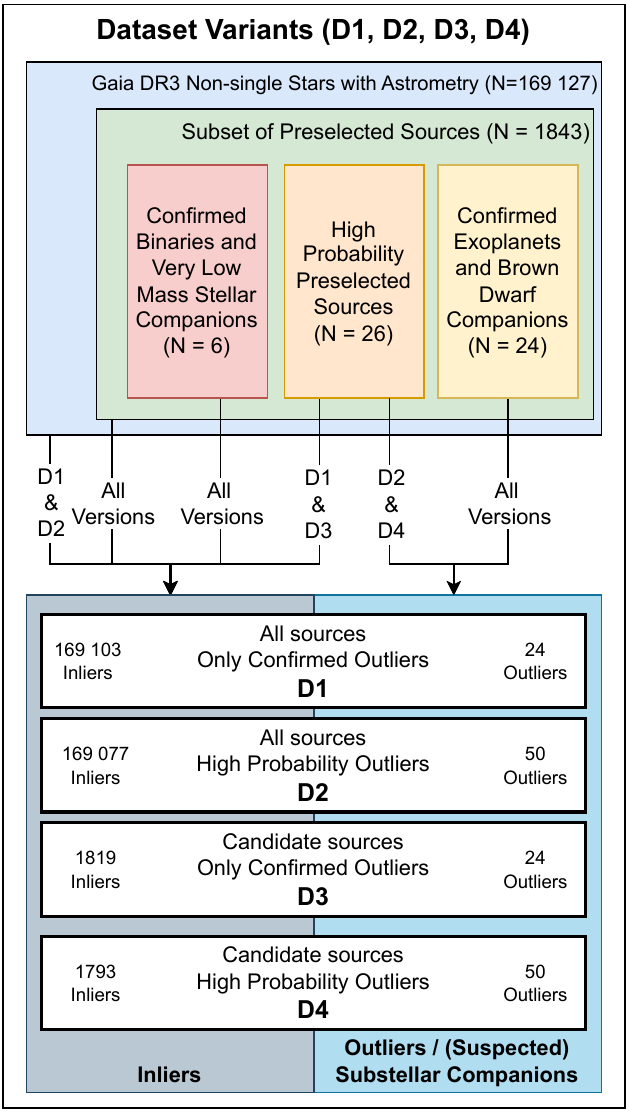}
\caption{Overview of the data and different partitionings (Datasets D1, D2, D3, D4) used. Source preselection was performed as described in Section \ref{label_definition}.}
\label{fig:dataset}
\end{figure}

\section{\gaia DR3 Data selection and augmentation}
The principle data source for this study are two tables in the \gaia DR3 archive\footnote{\url{https://gea.esac.esa.int/archive/}}: first the main astrometric catalog published in a table named \texttt{gaiadr3.gaia\_source} resulting from fitting a single-star model, and second the Keplerian-orbit solutions recorded in \texttt{gaiadr3.nss\_two\_body\_orbit}, which were obtained by fitting a non-single star (NSS) model.
We queried the \texttt{nss\_two\_body\_orbit} table and selected all astrometric orbits, i.e.\ entries having a \nssSolutionType of either \typeOrb, \typeASB, \typeOrbAltStar, or \typeOrbTarStar\footnote{{\typeOrbAltStar is a shorthand for either \typeOrbAlt or \typeOrbAltVal and \typeOrbTarStar is a shorthand for \typeOrbTar or \typeOrbTarVal.}}. The former two types were produced by the `binary pipeline` \citep{2023A&A...674A...9H}, whereas \typeOrbAltStar and \typeOrbTarStar solutions were produced by the `exoplanet pipeline` \citep{2023AA...674A..10H}. The exoplanet pipeline produced 198 solutions of type \typeOrbTarVal, which correspond to systems whose orbits where confirmed with independent datasets \citep{2023AA...674A..10H}.

For 98 sources that have both \typeASB and \typeOrbTarStar solutions 
we only retained the \typeASB solutions. Two solutions were initially identified as false positives by the \gaia collaboration\footnote{\url{https://www.cosmos.esa.int/web/gaia/dr3-known-issues\#FalsePositive}} and we discarded those. Two more false-positive solutions were reported during the refereeing process, which we did not account for here but they are discussed in Section \ref{sec:epcandidates}. This left us with 169\,127 orbit solutions for the same number of sources.

For these sources, we retrieved the data of the \gstable table which contains magnitudes, colours, radial velocities, and other information obtained from the astrometric single-star solution \citep{Lindegren:2021vn}. 
The \gaia DR3 table fields are described in the online documentation \citep{2022gdr3.reptE....V}, but not all are relevant for the purpose of our study. The selection of relevant fields and the augmentation with additional fields is described below.

The \gstable and \nssTBO fields that we used are listed in Tables \ref{tab:gs_columns} and \ref{tab:tbo_columns}, respectively. We augmented these tables with a few fields to make relevant source or orbit parameters directly accessible and interpretable. For the \texttt{gaia\_source} table, we computed the absolute source $G$-band magnitude from the apparent magnitude \texttt{phot\_g\_mean\_mag} and the single-star model parallax $\varpi_\mathrm{ss}$ from \gstable without zero-point correction \citep{2021A&A...649A...4L}.

For the \nssTBO table, we converted the Thiele-Innes coefficients ($A,B,F,G$) into the geometric orbital elements (the semimajor axis of the photocentre orbit $a_0$, the argument of periastron $\omega$, the longitude of the ascending node $\Omega$, and the orbital inclination $i$) using the linearised formulae described in \citet{2023A&A...674A...9H,2023AA...674A..34G,2023AA...674A..10H} and taking into account all parameter covariances using code implemented in \texttt{pystrometry} \citep{johannes_sahlmann_2019_3515526}. We also computed the astrometric mass function in solar masses as described in \citet[][Eq.\ 13]{2023A&A...674A...9H}:
\begin{equation} \label{eq:fm}
    f_M = 365.25^2 \frac{a_0^3}{P^2 \varpi_\mathrm{nss}^3}, 
\end{equation}
where $P$ is the orbital period in days and $\varpi_\mathrm{nss}$ is the non-single-star model parallax from \nssTBO in mas. Finally we joined both tables on the source identifier \texttt{source\_id}.

We note that the \nssSolutionType is disregarded in the machine learning analysis. This is intentional because we want to apply our approach to purely astrometric orbits with as little selection effects as possible. A \nssSolutionType = \typeASB, for instance, implies that a compatible orbit was detected in two \gaia instruments at the same time, imposing additional constraints on source apparent magnitude, and the necessary RV amplitude makes it unlikely that an exoplanet signature corresponds to a \typeASB.

\subsection{Data imputation}\label{sec:imputation}
For some sources/solutions not all table fields have been filled, e.g.\ sources that are too faint to be observed with the \gaia Radial Velocity Spectrometer \citep[RVS,][]{2023A&A...674A...5K} do not have an entry in the \texttt{radial\_velocity} field. We therefore filled these occurrences with appropriate numerical values. The affected fields and fill-values are listed in Table \ref{tab:fillvalues}. The fill values were determined as the median of the remaining population, except for radial\_velocity and eccentricity\_error where they were set to zero\footnote{The median radial\_velocity of sources where it is available is $-2.48$ km/s, thus generally the sample average value is compatible with zero.}.

\begin{table}
\centering
\caption{Fill values used for data imputation. The table \texttt{nss\_tbo} stands for \nssTBO, FV stands for \emph{fill value}, and $N$ indicates the number of values filled.} \label{tab:fillvalues}
\begin{tabular}{rrrrr}
\hline
Table & Field & FV &Unit& $N$\\
\hline
\gstable & radial\_velocity & 0 & km/s & 21460\\
\gstable & radial\_velocity\_error&2.41 & km/s & 21460\\
\gstable & phot\_rp\_mean\_mag & 12.67 & mag & 2\\
\gstable & phot\_bp\_mean\_mag & 13.70 & mag & 2\\
\gstable & bp\_rp & 0.99 & mag & 2\\
\gstable & phot\_bp\_rp\_excess\_factor &  1.22 & & 2\\
\texttt{nss\_tbo} & eccentricity\_error & 0 & &1979\\
\hline
\end{tabular}
\end{table}

\subsection{Definition of labels}\label{label_definition}
To assign labels to individual sources and orbital solutions, we used the \nssBM table \citep[][Sect.\ 5]{2023AA...674A..34G} and the literature. 

131\,142 of our selected sources have entries in \nssBM. 105 of those have several mass estimates {originating from different mass estimation methods as indicated by the field} \nssBM.\texttt{combination\_method}. We removed {entries listed with a combination method that includes `SB2' and `Eclipsing', i.e.\ that rely on a double-lined spectroscopic or eclipsing binary solution, respectively,} which left us with 131\,037 unique sources. The \sscc label was assigned to 1838 sources with \nssBM.\texttt{m2\_lower} $<0.08$ and the \bsscc label was assigned to sources with \nssBM.\texttt{m2\_upper} $<0.08$ \citep[cf.][]{2023MNRAS.526.5155S}. In the context, the \sscc label refers to solutions that correspond to sources, which based on their entries in \nssBM could be substellar-companion candidates.

Finally, we retrieved from the literature all the solutions that correspond to confirmed exoplanets (companion mass $M_2 \lesssim 20 M_\mathrm{Jup}$), brown-dwarf companions ($20 M_\mathrm{Jup} \lesssim M_2 \lesssim 80 M_\mathrm{Jup}$), and binary stars ($0.12 M_{\sun} \lesssim M_2$), where within that last category we also marked very-low-mass stellar companions with $80 M_\mathrm{Jup} \lesssim M_2 \lesssim 0.12 M_{\sun}$. Table \ref{tab:labelcount} shows the label counts and Table \ref{tab:labeldef} lists the individual solutions, their labels, and the respective references.

\begin{table}
\centering
\caption{Label counts for our dataset.} \label{tab:labelcount}
\begin{tabular}{ll}
\hline
Label & Count \\
\hline
\texttt{\sscc} & 1787 \\
\texttt{\bsscc} & 26 \\
\bdc & 14 \\
\exoplanet & 10 \\
\vlmsc & 3 \\
\bs & 3 \\
\fpo & 2 \\
\hline
\end{tabular}
\end{table}

\section{Data analysis}
\begin{figure} 
\centering
\includegraphics[width=\linewidth]{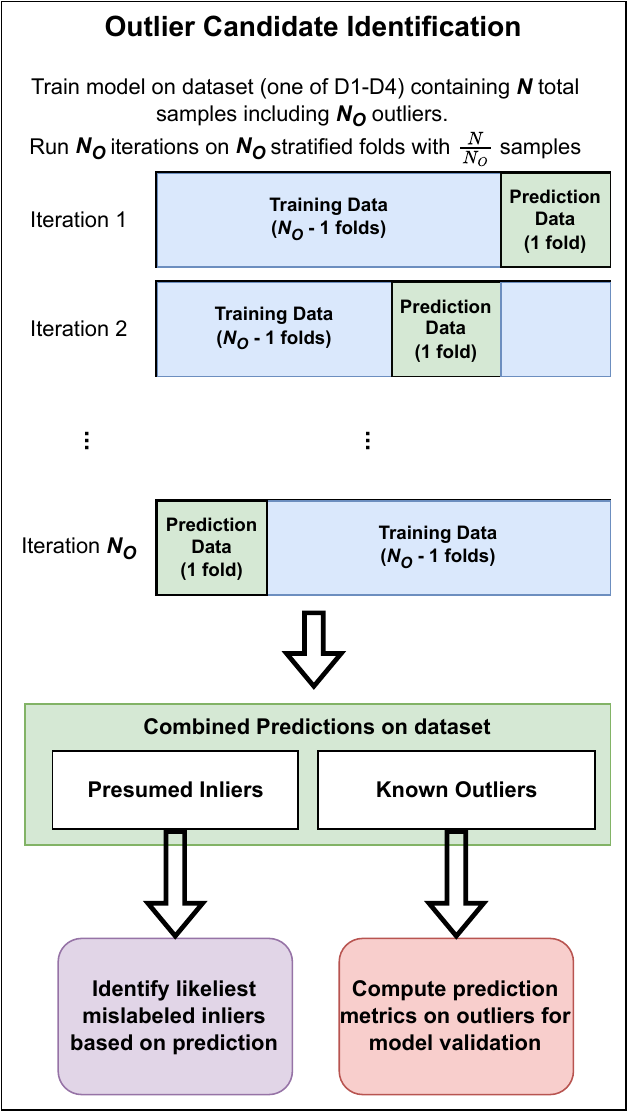}
\caption{Overview of the system architecture used for identifying candidates.} \label{fig:prediction}
\end{figure}

\subsection{Substellar Companions as Outliers}
Fundamentally, we expect the number of substellar companions in the non-single sources to be quite small, in the range of less than one percent and in the single digit percentile range amongst the \sscc. As such, we can consider them to be outliers in our data and we expect them to be identifiable based on their \gaia DR3 parameters.

Even though some relations between the astrometric data, such as the value of the mass function $f_M$ (see Equation \ref{eq:fm}), can be strongly indicative of the potential presence of a substellar companion, it is conceivable that there are various correlations and interactions between different parameters and the likelihood of a specific source having a substellar companion can likely not easily be reduced to a single parameter. Conversely, there are likely also other forms of outliers in the data, such as anomalies in the data processing pipeline or very low-mass stellar companions. Thus, we formulate the problem of identifying likely exoplanets and brown-dwarf companions as two-fold: first, in identifying outliers in the dataset and, secondly, identifying those outliers that are particularly likely to be substellar companions. We achieve this by formulating a semi-supervised outlier detection problem, in which the outliers used in training are previously identified substellar companions and not other outliers such as data processing anomalies or very low-mass stellar companions.

\subsection{Semi-supervised Outlier Detection}
Given the availability of some but few labelled examples (24 confirmed substellar companions), we choose a semi-supervised machine learning approach, i.e.\ one that relies on a small number of labelled samples and a larger number of unlabelled samples during training.
Further, given the difficulty of solving the task at all due to the limited amount of confirmed samples  we consider different dataset partitionings in the training and ensemble across configurations as later explained in this section. 

In particular, as shown in Figure \ref{fig:dataset}, we build four dataset versions (D1, D2, D3, D4), which contain subsets of the data (e.g.\ only \sscc in D3 and D4) and add 26 \bsscc as positively labelled outliers (D2 and D4). 

In most established machine learning approaches \citep{zhao2018xgbod,han2022adbench}, the task of semi-supervised outlier detection is approached by training on the limited amount of data containing both known outliers and inliers and then predicting on a separate dataset. Since we only have a few confirmed outliers and virtually no confirmed inliers (i.e. binaries and very low mass stellar companions) we developed a different approach and instead consider our entire dataset of inliers (i.e. the vast majority of the data) to be weakly labelled, i.e. potentially mislabelled. Thus, we transform the task from classifying newly unseen samples into inliers or outliers into identifying the likeliest mislabelled samples in our training data. To this end, we perform a k-fold cross-validation and use the predictions on the presumed inliers as the indication of whether we believe that sample to be a mislabelled one. This procedure is shown in detail in Figure \ref{fig:prediction}. We choose the amount of folds to be identical to the number of outliers to maximise the amount of confirmed outliers in the training data. The left-out outlier in the prediction data fold is used to compute the validation metrics in the results section.

To further improve the robustness of the approach and since precise performance evaluation is challenging with the limited amount of data
, we use two different approaches in our ensembles. In particular, we utilise XGBoost \citep{chen2016xgboost} as implemented in the Python package of the same name\footnote{\url{https://xgboost.readthedocs.io/en/stable/index.html} Accessed: 2024-03-08} and a balanced random forest classifier \citep{chen2004using} as implemented in the \textit{imblearn}\footnote{\url{https://imbalanced-learn.org/} Accessed: 2024-03-08} Python module. Given the invariance of both approaches to monotonous transforms, no data transforms were applied beyond the elimination of samples (i.e. sources/orbital solutions) for which any of the studied features was not available.

\subsection{Feature Selection and Importance}\label{sec:features}

The original data contain a total of 47 unique data points per source (an overview of all parameters is given in Tables \ref{tab:gs_columns} and \ref{tab:tbo_columns}). During the development of the approach we used different feature-importance measures such as SHapley Additive exPlanations (SHAP) values \citep{lundberg2017unified} and the feature importance implemented in \textit{scikit-learn} \citep{pedregosa2011scikit}. Using this we identified and discarded several confounders, i.e.\ features that are marked as important for identifying outliers but that are not expected as such based on physical arguments.  

As examples, we discarded both the parallax and the apparent magnitude of the source. These correlate positively with having confirmed substellar companions because those are the best-surveyed sources with ground-based, high-precision radial-velocity observations and generally are easier to follow-up. From an astrophysical standpoint, there is little reason for a bright source to be more likely to host a substellar companion than a faint one, at least within the parameter space surveyed here. The other confounders that we identified are indicated in Tables \ref{tab:gs_columns} and \ref{tab:tbo_columns}.

Additionally, we use SHAP values as provided by the Python module\footnote{\url{https://shap.readthedocs.io/en/latest/} Accessed: 2024-03-08} of the same name to study the feature importance amongst our proposed candidates as well as to identify mislabelled training samples as shown in Section \ref{sec:results}.

\subsection{Ensembling and Candidate Selection}\label{sec:cand_selection}

As outlined in the previous paragraphs, we use four different datasets (see Figure \ref{fig:dataset}) and two different models (XGBoost and BalancedRandomForest). For each of the eight possible dataset and model combinations we compute a prediction for each source in a binary classification with value 0 to 1, where higher values mean the source is likelier to be an outlier according to the model using the scheme described in Figure \ref{fig:prediction}.  

These scores are not probabilities and are not calibrated, hence we cannot directly combine the output of different model and dataset combinations. Instead, we build our predictions using max-voting on the highest scoring top 50 samples per configuration. Only samples that are classified as presumed inliers are considered in those top 50. From these top 50 samples per configuration, we take those samples that occur most often across the different configurations and describe this relative occurrence rate by $\rho$ (0 if in none to 1 if in all). Note that not all samples are in all datasets and individual samples may in total occur as inliers zero (confirmed substellar companions - always outliers), two (\bsscc, orange in Figure \ref{fig:dataset}), four (non-single sources) or eight (\sscc) times. In addition to the overall relative occurrence $\rho$, which describes the rate of occurrence in the eight combinations, we observe occurrence rates $\rho_{nss}$ and $\rho_{ssc}$, which are the rates of occurrence in the combinations using large datasets (D1 \& D2 containing non-single stars) and smaller ones (D3 \& D4 containing only preselected sources), respectively. Details on the datasets are displayed in Figure \ref{fig:dataset}.

\section{Results: substellar-companion candidates}\label{sec:results}
\begin{figure}
    \centering
    \includegraphics[width=\columnwidth]{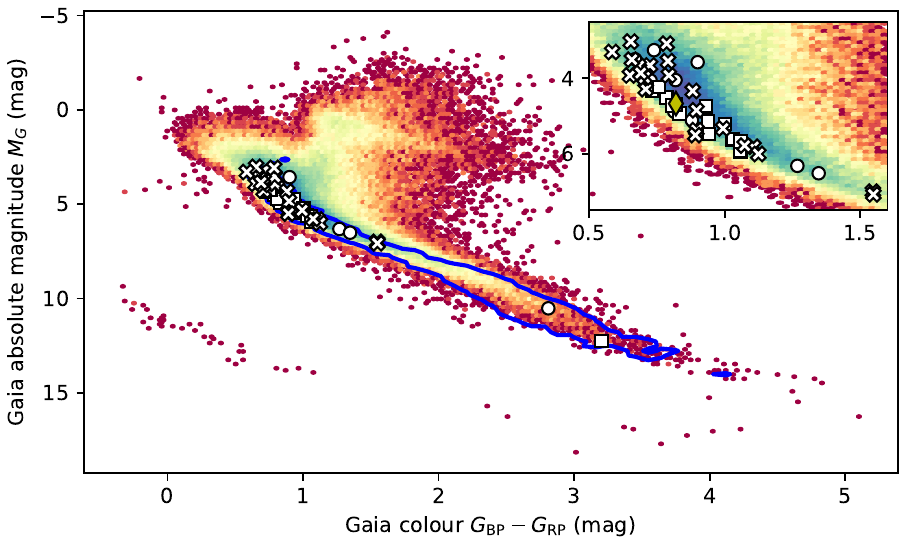}
    \includegraphics[width=\columnwidth]{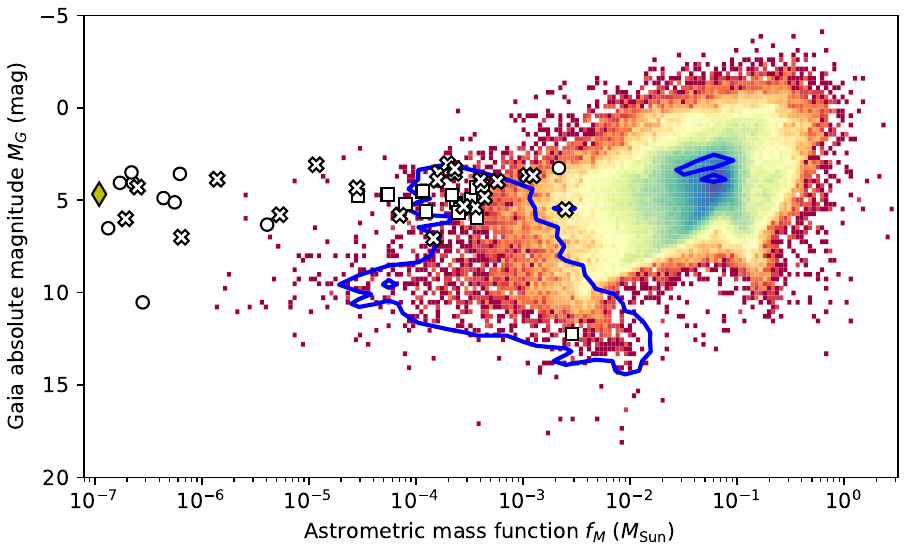}
    \caption{Density histograms of absolute magnitude as a function of colour (top panel, the inset show a zoom into the region of interest) and mass function (bottom) for all \gaia DR3 astrometric orbits. The blue contours indicate the concentration of the solutions labelled as \sscc. Circles indicate the 10 confirmed exoplanets, squares indicate the 14 confirmed BD-companions, and crosses indicate our 22 best candidates for substellar companions discussed in the text. For reference, the Sun has $M_G=4.67$ and $G_\mathrm{BP}-G_\mathrm{RP}=0.82$ \citep{2018MNRAS.479L.102C}, and the mass function of a $5\,M_J$ planet in a Jupiter-like orbit around the Sun is $1.1 \cdot 10^{-7}$. These reference locations are marked with a yellow diamond. The exoplanet with the largest mass function is HD\,39392\,b \citep{Wilson:2016aa,2023MNRAS.526.5155S}.}
    \label{fig:overview}
\end{figure}

We applied a cutoff in relative occurrence at $\rho>0.5$ which resulted in the identification of our 14 best candidates for exoplanets and brown-dwarf companions. In addition we identified four  candidates with $\rho_{nss}>0.5$ and four candidates with $\rho_{ssc}>0.5$. These three categories are presented in their respective sections in Table \ref{tab:candidates1}. All 22 candidates are shown in Figure \ref{fig:overview} and discussed below. We also identified lower-confidence candidates using  a lower cutoff value of $\rho>0.125$, which we list in Table \ref{tab:candidates2}.

\subsection{Validation}

Ensuring the performance of our approach is a challenging task given the small amount of labelled samples. In the prediction method described in Figure \ref{fig:prediction} we obtain the predictions on the left-out fold as in a stratified K-fold validation. We use these predictions to compute performance metrics. As the inlier labels in these folds are weak labels (we expect some substellar companions in them) the metrics are probably underestimating the true performance since the models correctly "mislabel" some false inliers. Similarly, for the high probability (but not confirmed) outliers some may be mislabelled outliers, further impacting the performance measurements. Detailed results for different configurations are given in Table \ref{table:validation}. Overall, we achieve very good results on the area under the receiver operating characteristic (AUROC) and accuracy of predictions. The achieved area under the precision recall curve (AUPRC) results are further from the optimal value of 1 but this is expected given the weak labels (in some cases the model may be right but the label wrong) and challenging task due to the significant class imbalance (less than 0.03\% labelled outliers amongst all sources).

\begin{table*}
\centering
\caption{Validation results for different datasets and models (Random Forest - RF, XGBoost - XGB) showing accuracy of prediction as well as area under the receiver operating characteristic (AUROC) and area under the precision recall curve (AUPRC). {Results are averaged over the relevant runs, e.g. RF on D1, RF on D3, XGB on D1, XGB on D3 for the first column.}} 
\label{table:validation}
\begin{tabular}{cccccccc}
\hline
\makecell{Datasets \\ \& Models \\ \hline Metric} & \makecell{Only Confirmed Outliers \\D1\&D3 \\RF+XGB} & \makecell{High Probability Outliers \\ D2\&D4 \\RF+XGB} & \makecell{All Sources \\D1\&D2 \\ RF+XGB} & \makecell{Candidate Sources \\D3\&D4 \\RF+XGB} & \makecell{All Datasets \\D1-D4\\ XGB}& \makecell{All Datasets \\D1-D4\\ RF}& All \\
\hline
Accuracy & 0.97 & 0.94 & 0.98 & 0.96 & 0.99 & 0.92 & 0.96 \\
AUROC & 0.98 & 0.97 & 0.99 & 0.97 & 0.98 & 0.97 & 0.97 \\
AUPRC & 0.65 & 0.49 & 0.5 & 0.57 & 0.66 & 0.47 & 0.56 \\
\hline
\end{tabular}
\end{table*}

A secondary measure of the performance of our approach is the accuracy of the applied majority voting scheme to identify the best candidates from the eight configurations. A total of 170 unique sources show up in the top 50 outliers of the eight approaches, i.e.\ amongst all samples including presumed inliers and outliers. Among those 170, we find 22 of our 24 confirmed substellar companions among the 61 highest ranked samples across all configurations, each being identified in at least two separate runs. The only missing ones are the brown dwarf companions in LHS 1610 \citep[\gaia DR3 43574131143039104,][]{2024AJ....168..140F} and HD 89707 \citep[\gaia DR3 3751763647996317056,][]{2023AA...674A..34G}. The one around LHS 1610 was likely missed due to the very-red colour of its M5-dwarf host star, which is uncommon in the confirmed substellar companions. The one around HD 89707 was barely below the detection threshold scoring highly on most models. We investigate this further in Section \ref{sec:featureImportance} on feature importance. 
\begin{table*}
\centering
\caption{Best candidates for exoplanets and brown-dwarf companions. The SpT column indicates the source's spectral type as listed in Simbad, the solution type column indicates a short-hand for the \nssSolutionType (similar to \citealt{2023AJ....165..266M}: OTSV is \typeOrbTarVal, OTS is \typeOrbTar, and ASB1 is \typeASB), $M_2$ is a companion-mass estimate under the dark-companion assumption and using the primary-mass estimate $M_1$ from \nssBM, $\rho$ is the relative occurrence introduced in Section \ref{sec:cand_selection}, $\rho_\mathrm{nss}$ is the relative occurrence for dataset variants D1\&D2 only, $\rho_\mathrm{ssc}$ is the relative occurrence for dataset variants D3\&D4 only, the $ssc$ flag is 1 when the solution was part of `\sscc` or `\bsscc` and 0 otherwise, and $M_{2, \mathrm{alt}}$ is the companion-mass range as estimated in \nssBM. The orbital solutions of {* 54 Cas} and {BD+75 510} were retracted as false positives by the Gaia collaboration.}
\label{tab:candidates1}
\footnotesize    
\begin{tabular}{rrrrrrrrrrrrr}
\hline
\gaia DR3 source\_id & Name & SpT & sol.  & $P$ & $f_M$ & $M_2$ & $\rho$ & $\rho_\mathrm{ssc}$ & $\rho_\mathrm{nss}$ & $ssc$ & $M_1$ & $M_{2, \mathrm{alt}}$\\
& & &type &(day) &($M_{\sun}$) & ($M_\mathrm{Jup}$) & & & & & ($M_{\sun}$) & ($M_{\sun}$)\\
\hline
2884087104955208064 & \href{http://simbad.u-strasbg.fr/simbad/sim-basic?Ident=Gaia+DR3+2884087104955208064&submit=SIMBAD+search}{HD  40503} & K2V & OTSV & 826.5 & 1.9e-07 & 5.2 & 0.8 & 0.8 & 0.8 & 1 & ${0.79}^{+0.05}_{-0.23}$ & $0.004$-$0.681$ \\
522135261462534528 & \href{http://simbad.u-strasbg.fr/simbad/sim-basic?Ident=Gaia+DR3+522135261462534528&submit=SIMBAD+search}{*  54 Cas} & F8 & OTS & 401.1 & 2.5e-07 & 7.1 & 0.9 & 1.0 & 0.8 & 1 & ${1.09}^{+0.06}_{-0.31}$ & $0.005$-$0.905$ \\
1712614124767394816 & \href{http://simbad.u-strasbg.fr/simbad/sim-basic?Ident=Gaia+DR3+1712614124767394816&submit=SIMBAD+search}{BD+75   510} & M0V & OTSV & 297.6 & 6.4e-07 & 7.3 & 0.8 & 0.8 & 0.8 & 1 & ${0.71}^{+0.05}_{-0.20}$ & $0.006$-$0.632$ \\
1610837178107032192 & \href{http://simbad.u-strasbg.fr/simbad/sim-basic?Ident=Gaia+DR3+1610837178107032192&submit=SIMBAD+search}{HD 128717} & F8 & OTS & 1089.2 & 1.4e-06 & 12.8 & 0.9 & 1.0 & 0.8 & 1 & ${1.13}^{+0.06}_{-0.16}$ & $0.010$-$1.109$ \\
1878822452815621120 & \href{http://simbad.u-strasbg.fr/simbad/sim-basic?Ident=Gaia+DR3+1878822452815621120&submit=SIMBAD+search}{BD+24  4592} & K2 & OTS & 1009.3 & 5.4e-06 & 16.3 & 0.9 & 1.0 & 0.8 & 1 & ${0.82}^{+0.05}_{-0.21}$ & $0.014$-$0.766$ \\
1897143408911208832 & \href{http://simbad.u-strasbg.fr/simbad/sim-basic?Ident=Gaia+DR3+1897143408911208832&submit=SIMBAD+search}{HD 207740} & G5V & OTS & 712.2 & 2.8e-05 & 32.3 & 0.9 & 1.0 & 0.8 & 1 & ${0.99}^{+0.05}_{-0.12}$ & $0.029$-$1.089$ \\
6330529666839726592 & \href{http://simbad.u-strasbg.fr/simbad/sim-basic?Ident=Gaia+DR3+6330529666839726592&submit=SIMBAD+search}{TYC 4998-437-1} & -- & Orbital & 526.1 & 7.1e-05 & 39.0 & 0.6 & 0.5 & 0.8 & 1 & ${0.82}^{+0.05}_{-0.20}$ & $0.032$-$0.831$ \\
364792020789523584 & \href{http://simbad.u-strasbg.fr/simbad/sim-basic?Ident=Gaia+DR3+364792020789523584&submit=SIMBAD+search}{BD+32    92} & F8 & OTS & 291.9 & 1.6e-04 & 67.2 & 1.0 & 1.0 & 1.0 & 1 & ${1.21}^{+0.06}_{-0.14}$ & $0.057$-$1.382$ \\
5773484949857279104 & \href{http://simbad.u-strasbg.fr/simbad/sim-basic?Ident=Gaia+DR3+5773484949857279104&submit=SIMBAD+search}{GSC 09436-01089} & -- & ASB1 & 409.9 & 2.3e-04 & 73.8 & 0.6 & 0.8 & 0.5 & 1 & ${1.16}^{+0.07}_{-0.12}$ & $0.063$-$1.391$ \\
3913728032959687424 & \href{http://simbad.u-strasbg.fr/simbad/sim-basic?Ident=Gaia+DR3+3913728032959687424&submit=SIMBAD+search}{HD  99251} & F5 & Orbital & 328.9 & 2.3e-04 & 78.6 & 0.9 & 1.0 & 0.8 & 1 & ${1.28}^{+0.06}_{-0.20}$ & $0.069$-$1.411$ \\
2540855308890440064 &  & N/A & Orbital & 239.8 & 2.8e-04 & 65.7 & 0.9 & 1.0 & 0.8 & 1 & ${0.88}^{+0.06}_{-0.17}$ & $0.056$-$1.006$ \\
1156378820136922880 & \href{http://simbad.u-strasbg.fr/simbad/sim-basic?Ident=Gaia+DR3+1156378820136922880&submit=SIMBAD+search}{G  15-6} & G4 & Orbital & 268.3 & 3.6e-04 & 72.2 & 0.8 & 1.0 & 0.5 & 1 & ${0.88}^{+0.05}_{-0.30}$ & $0.065$-$0.872$ \\
2171489736355655680 & \href{http://simbad.u-strasbg.fr/simbad/sim-basic?Ident=Gaia+DR3+2171489736355655680&submit=SIMBAD+search}{HD 206484} & F8 & ASB1 & 479.5 & 4.0e-04 & 90.4 & 0.8 & 1.0 & 0.5 & 1 & ${1.18}^{+0.06}_{-0.06}$ & ${0.078}^{+0.006}_{-0.006}$ \\
3909531609393458688 & \href{http://simbad.u-strasbg.fr/simbad/sim-basic?Ident=Gaia+DR3+3909531609393458688&submit=SIMBAD+search}{TYC  277-599-1} & -- & Orbital & 520.6 & 4.4e-04 & 79.5 & 0.8 & 1.0 & 0.5 & 1 & ${0.92}^{+0.06}_{-0.13}$ & $0.067$-$1.141$ \\
\hline
4545802186476906880 & \href{http://simbad.u-strasbg.fr/simbad/sim-basic?Ident=Gaia+DR3+4545802186476906880&submit=SIMBAD+search}{HD 153376} & F8V & OTS & 910.8 & 1.2e-05 & 24.2 & 0.4 & 0.0 & 0.8 & 0 & NaN & NaN \\
5148853253106611200 & \href{http://simbad.u-strasbg.fr/simbad/sim-basic?Ident=Gaia+DR3+5148853253106611200&submit=SIMBAD+search}{LP  769-9} & -- & Orbital & 339.6 & 1.5e-04 & 44.8 & 0.4 & 0.0 & 0.8 & 1 & ${0.69}^{+0.05}_{-0.19}$ & $0.039$-$0.723$ \\
3921176983720146560 & \href{http://simbad.u-strasbg.fr/simbad/sim-basic?Ident=Gaia+DR3+3921176983720146560&submit=SIMBAD+search}{HD 106888} & F8 & OTSV & 366.3 & 5.8e-04 & 98.9 & 0.5 & 0.0 & 1.0 & 0 & ${1.11}^{+0.06}_{-0.14}$ & $0.083$-$1.388$ \\
3067074530201582336 & \href{http://simbad.u-strasbg.fr/simbad/sim-basic?Ident=Gaia+DR3+3067074530201582336&submit=SIMBAD+search}{BD-06  2423A} & -- & OTS & 43.6 & 2.5e-03 & 155.8 & 0.4 & 0.0 & 0.8 & 0 & NaN & NaN \\
\hline
5484481960625470336 & \href{http://simbad.u-strasbg.fr/simbad/sim-basic?Ident=Gaia+DR3+5484481960625470336&submit=SIMBAD+search}{HD  49264} & G3V & Orbital & 428.0 & 1.6e-04 & 59.7 & 0.5 & 0.8 & 0.2 & 1 & ${1.02}^{+0.06}_{-0.08}$ & $0.052$-$1.261$ \\
2280560705703031552 & \href{http://simbad.u-strasbg.fr/simbad/sim-basic?Ident=Gaia+DR3+2280560705703031552&submit=SIMBAD+search}{HD 212620} & F8 & Orbital & 429.0 & 2.0e-04 & 76.1 & 0.4 & 0.8 & 0.0 & 1 & ${1.32}^{+0.06}_{-0.19}$ & $0.065$-$1.475$ \\
5323844651848467968 & \href{http://simbad.u-strasbg.fr/simbad/sim-basic?Ident=Gaia+DR3+5323844651848467968&submit=SIMBAD+search}{HD  78631} & F8V & ASB1 & 666.5 & 1.1e-03 & 123.6 & 0.5 & 1.0 & 0.0 & 1 & ${1.11}^{+0.06}_{-0.06}$ & ${0.076}^{+0.014}_{-0.012}$ \\
1576108450508750208 & \href{http://simbad.u-strasbg.fr/simbad/sim-basic?Ident=Gaia+DR3+1576108450508750208&submit=SIMBAD+search}{HD 104289} & F8 & OTS & 1233.3 & 1.2e-03 & 132.7 & 0.5 & 1.0 & 0.0 & 1 & ${1.15}^{+0.06}_{-0.05}$ & ${0.047}^{+0.006}_{-0.006}$ \\
\hline
\end{tabular}
\end{table*}

\subsection{Exoplanet candidates}\label{sec:epcandidates}
Table \ref{tab:candidates1} is ordered by increasing mass function and we will discuss the individual systems here. 
We identified five high-probability exoplanet candidates. All of them are previously-known exoplanet candidates and have \gaia astrometric object of interest identifiers (Gaia-ASOI-\#\footnote{\url{https://www.cosmos.esa.int/web/gaia/exoplanets}}):

The companion of HD\,40503 \citep[Gaia-ASOI-002,][]{2023AA...674A..34G,2023AA...674A..10H,2023AJ....165..266M,2022AJ....164..196W} continues to await full RV confirmation. 

We identify HD\,128717 (Gaia-ASOI-009) as a strong candidate for hosting a super-Jupiter, motivating prioritised follow-up of this star. 

The companion of *\,54\,Cas (Gaia-ASOI-003) is an exoplanet candidate discussed in \citet{2023AA...674A..34G}, however, the corresponding orbital solution was retracted as false positive by the \gaia collaboration in May 2024\footnote{\url{https://www.cosmos.esa.int/web/gaia/dr3-known-issues\#FalsePositive}}, thus this is no longer an exoplanet candidate. 

The star BD$+$75\,510 \citep[HIP\,66071, Gaia-ASOI-001,][]{2023AA...674A..10H, 2023AA...674A..34G} was followed up with high-precision RVs and a sub-Jupiter mass exoplanet was discovered \citep{2023A&A...677L..15S}. However, also this orbital solution was retracted as false positive by the \gaia collaboration in May 2024. The exoplanet discovered by \citep{2023A&A...677L..15S} does therefore not have a corresponding orbit solution in \gaia DR3 and this is no longer an exoplanet candidate in the context of our study.

The companion of BD$+$24\,4592 (Gaia-ASOI-005) appears to be a strong exoplanet candidate in our analysis on the basis of the \gaia DR3 orbit, but this star was previously identified as spectroscopic binary \citep[G\,127$-$33,][]{Latham:2002zr} with a period of 3977 days and it is listed in the SB9 catalog \citep{Pourbaix:2004yq} with a large RV amplitude of 2.7 km/s. The \gaia astrometric period of 1009 days is very close to the \gaia DR3 time range. It is therefore plausible that the \gaia orbit is spurious and the partially-covered binary-star orbit was interpreted as a low-amplitude orbit at $\sim$1/4 of the true period. We therefore argue that Gaia-ASOI-005 should no longer be classified as exoplanet candidate.

\subsection{Brown-dwarf companion candidates}
We identified 17 brown-dwarf companion candidates with mass estimates spanning the brown-dwarf desert but predominantly located on the heavy side ($\gtrsim60\,M_J$).

We identified the stars TYC\,4998-437-1, BD+32\,92, TYC\,277-599-1, LP\,769-9, HD\,49264, HD\,212620 as high-probability candidates for hosting brown-dwarf companions. For these six sources we did not find auxiliary information in the literature.

{We also identified HD\,207740 as host of a high-probability} brown-dwarf companion candidate. Three Keck HIRES spectra of this source are discussed in \citet{2020ApJ...898..119R} but no RVs appear to have been published.

The stars HD\,206484 and HD\,78631 have \gaia DR3 solution types of \typeASB which means that the \gaia RVS also detected a compatible RV variation. The companion mass estimates from \nssBM in these cases are therefore more reliable and straddle the substellar boundary, i.e.\ these companions are either high-mass brown dwarfs or very-low-mass stars. For HD\,78631 we noted a strong discrepancy between the companion-mass estimate from \nssBM of $\sim$80 \mj, which was determined using a non-zero flux ratio \citep{2023AA...674A..34G} and our estimate of $\sim$123 \mj. The latter was derived under the wrong assumption of a dark companion, and {a discrepancy can therefore 
 be expected}.

The star HD\,99251 hosts a brown-dwarf companion candidate. This star is also {EPIC 201928968}, a \textit{Kepler} eclipsing binary candidate with orbital period $<$1 day. No close companion was found with Speckle observations \citep{2016AJ....151..159S}.

The source \gaia DR3 2540855308890440064 hosts a brown-dwarf companion candidate. This source\_id is not resolved by Simbad and the source is a 10\arcsec\ visual binary with \href{http://simbad.u-strasbg.fr/simbad/sim-basic?Ident=TYC+4672-237-1&submit=SIMBAD+search}{{TYC 4672-237-1}}.

Our analysis of the \gaia DR3 \typeOrbTar orbit of HD\,104289 indicates a very low-mass stellar companion in a $1233\pm164$ day orbit. The true period, however, is almost twice as long with $\sim$2389 days as determined by ground-based RVs \citep{2019A&A...631A.125K}, who also detected the star's orbit in \hip astrometry at 2$\sigma$ confidence. 
Furthermore, this source has an independent \typeSB solution in the \nssTBO table with an even shorter period of $886\pm27$\,d, i.e.\ about one quarter of the true period. The companion-mass estimate from \nssBM listed in Table \ref{tab:candidates1} was determined from the combination of those discrepant solutions\footnote{The combination for mass estimation requires period agreement within $5\sigma$ \citep{2023AA...674A..34G}. An attempt for a combined astrometric-spectroscopic solution, however, is only made when the period values also agree within 10\,\%, see \url{https://gea.esac.esa.int/archive/documentation/GDR3/Data_analysis/chap_cu4nss/sec_cu4nss_combined/ssec_cu4nss_combined_input.html}}, resulting in a significantly negative flux ratio and hence rendering these mass estimates unreliable\footnote{\url{https://www.cosmos.esa.int/web/gaia/faqs}}.
As shown by \citet{2019A&A...631A.125K}, the companion is indeed a very low-mass star and this is the second example in this work where the \gaia DR3 orbit is spurious because the binary's motion had not yet been monitored long enough.

GSC\,09436-01089 has a solution type of \typeASB which means that the \gaia RVS also detected a compatible RV variation. Using the primary-mass estimate of ${M_1} = 1.16 \pm 0.12\,M_{\sun}$ from \nssBM and resampling the \gaia DR3 orbit parameters while accounting for all covariances\footnote{\label{covmat}{The correlation matrix of the \gaia DR3 orbital parameters is available in the archive and should be taken into account when propagating uncertainties to derived quantities, see e.g.\ \citet[][Appendix D.3]{2023AA...674A..34G}.}} we obtain a companion-mass estimate of ${M_2} = {74.88}^{+9.20}_{-8.07}\,{M_\mathrm{Jup}}$. GSC\,09436-01089 B is therefore likely a high-mass brown dwarf but it may also be a very-low-mass star.

Our analysis of the HD\,153376 astrometric orbit suggests that it is the likely host of a low-mass brown dwarf in a 910 day period. However, an extremely eccentric ($e\simeq0.8$) radial-velocity orbit of the same star was determined by \citet{2019A&A...631A.125K} with a period of $\sim$4900 days and a minimum mass of $\sim$0.24$M_{\sun}$. Similarly to BD+24\,4592 and HD\,104289, this is the third case of a long-period binary orbit being mistaken for a shorter-period and small-amplitude signal by the \gaia data processing.

The astrometric orbit of HD\,106888 has a compatible RV solution in terms of period and companion mass determined by \citet{2019A&A...631A.125K}. This had already been realised by the \gaia team in charge of the relevant \gaia DR3 processing \citep{2023AA...674A..10H} who assigned this solution the type \typeOrbTarVal. The companion mass is estimated at $\sim$0.1$M_{\sun}$, thus slightly above the substellar limit, yet our models have picked it up as an outlier in the D1\&D2 runs with $\rho_{nss}=1.0$. This companion has been detected directly \citep[e.g.][]{2020AJ....160....7T}. For completeness, HD\,106888 also has an outer, ultracool L1-dwarf companion at $\sim$38\arcsec\ separation \citep{2017MNRAS.470.4885M}, which makes this a triple system.

Our models in the D1\&D2 runs identified the orbit of BD-06\,2423A as indicative of a substellar companion, however, the companion mass is estimated at $\sim$0.15$M_{\sun}$, i.e.\ substantially above the substellar limit. This is also the candidate with by far the shortest orbital period of $\sim$44 days. This star is also a wide binary with an entry in the Washington Double Star Catalogue \citep{Mason:2001fk}, thus, this constitutes another triple system.

Finally, we present {G\,15-6} as the first independently-validated substellar companion identified by our work. The \gaia DR3 orbit implies a massive BD companion in a 268 day orbit. This source had already been identified as a binary from RV monitoring by \citet{Latham:2002zr} and it is listed in the SB9 catalogue with a period of 266 days. The minimum companion mass determined by \citet{Latham:2002zr} lies in the substellar range but the companion could not be confirmed as a brown dwarf because of the unknown orbital inclination. The \gaia astrometric orbit has now resolved this ambiguity. The matching periods and companion-mass estimates leave little doubt that the same orbital motion was detected independently by both \gaia and in the RVs from \citet{Latham:2002zr}.
Using the primary mass estimate of ${M_1} = 0.67\,M_{\sun}$ from \citet{Latham:2002zr} with a 10~\% uncertainty, and resampling the \gaia DR3 orbit parameters while accounting for all covariances\footref{covmat} we obtain a companion-mass estimate of ${M_2} = {62.43}^{+6.16}_{-5.44}\,{M_\mathrm{Jup}}$.
This companion therefore lies firmly in the substellar range and is the first confirmed brown-dwarf companion that was identified with our machine learning approach.

\subsection{Feature Importance Results}\label{sec:featureImportance}
In {this} section, we study the model predictions using SHAP values \citep{lundberg2017unified} to identify both the reasons for the models' predictions on the entire sample population and our candidate pool, but also for individual samples such as the confirmed brown dwarf the models missed.

\subsubsection{Discussion of selected features}\label{sec:selected_features}
It is helpful to inspect the information content of a few features in more detail:

Mass function ($f_M$): A small mass function is a necessary condition for a system with a substellar companion\footnote{We do not discuss brown-dwarf binaries here.}, but it is not a sufficient condition because twin-binary stars also have very small astrometric mass functions.

Radial-velocity error: As explained in the \href{https://gea.esac.esa.int/archive/documentation/GDR3/Gaia_archive/chap_datamodel/sec_dm_main_source_catalogue/ssec_dm_gaia_source.html#gaia_source-radial_velocity_error}{online documentation} and \citet{2023A&A...674A...5K,2023A&A...674A..32B}, the RV error given in \gstable has a stellar magnitude-dependent meaning. For bright sources (phot\_g\_mean\_mag $\lesssim$13) it is derived from the uncertainty of the median of the RV timeseries. In that case, a large RV error can therefore stem from large individual RV uncertainties or from unmodelled RV variations, e.g.\ as present for a binary star. For faint sources (phot\_g\_mean\_mag $\gtrsim$13), the RV error is determined by stacking the individual cross-correlation functions and therefore the sensitivity to binary-induced motion is different \citep[cf.][]{2022MNRAS.516.3661A}. In the sample of $\sim$ 170\,000 astrometric orbits with available RV error, about 43 \% fall in the `bright` category, and this percentage decreases to 36 \% for the \sscc. In addition to the bi-modal determination of the RV error, there is the general dependency on apparent stellar magnitude: the brighter the star, the higher the signal-to-noise in the spectra, and the smaller the RV error. As a result of these multiple dependencies it is often not directly interpretable how the RV error has influenced the models' predictions.

Goodness-of-fit: The goodness\_of\_fit from \nssTBO corresponds to the `gaussianized chi-square` statistic. For good orbital fits it should approximately follow a normal distribution with zero mean value and unit standard deviation. Values larger than roughly +3 indicate a bad fit. Our selection on astrometric orbits contains a mix of purely-astrometric orbits and combined RV+astrometric orbits, i.e.\ of type \typeASB. For the latter, the goodness\_of\_fit refers to the combined fit, hence we can expect different feature characteristics between both groups.

Flux excess: the phot\_bp\_rp\_excess\_factor \citep{2021A&A...649A...3R} contains a measure of the consistency between the red and blue source fluxes measured by \gaia. It has been shown that this metric can be sensitive to near-equal-brightness ($\Delta$mag $<0.5$) binaries \citep{2020MNRAS.496.1922B}.

Colour and absolute magnitude: Finally, the \gaia colour $G_\mathrm{BP}-G_\mathrm{RP}$ ($\equiv$ bp\_rp) and absolute magnitude $M_G$ ($\equiv$ absolute\_phot\_g\_mean\_mag) indicate the location of the source in the colour-magnitude diagram (Fig.\ \ref{fig:overview} top).

\subsubsection{Predictions for populations}
\begin{figure} 
\centering
\includegraphics[width=\linewidth]{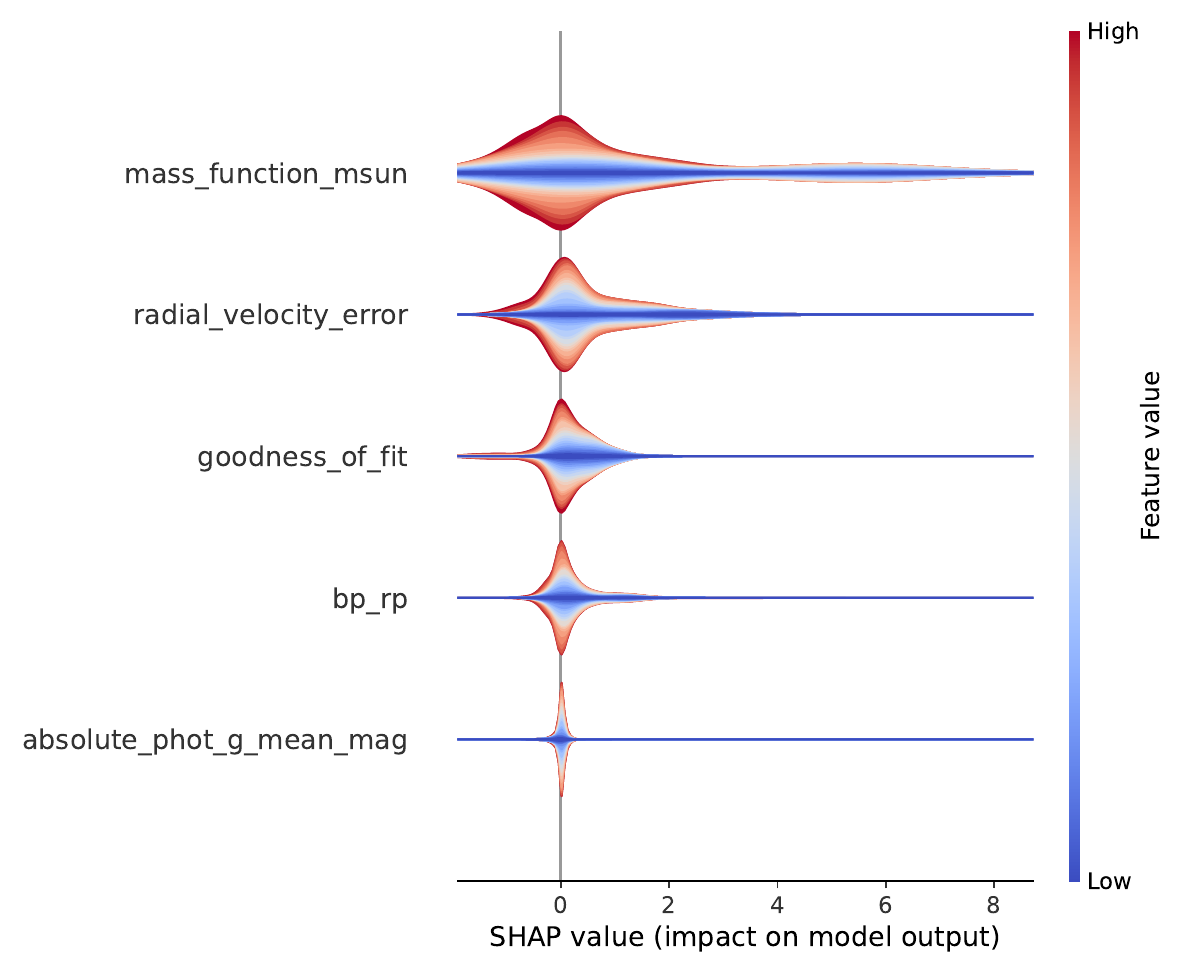}
\caption{SHAP value distribution for the top five features, {i.e.\ the input parameters such as radial velocity error}, of the top 50 sources identified in each of the 8 configurations (total number of 8$\times$50 sources); colour indicates distribution of the feature; especially smaller values of the variables are associated with higher SHAP values. {The top five features are determined as having the highest average over the entire sample.}\label{fig:candidate_shap}}
\end{figure}

\begin{figure} 
\centering
\includegraphics[width=\linewidth]{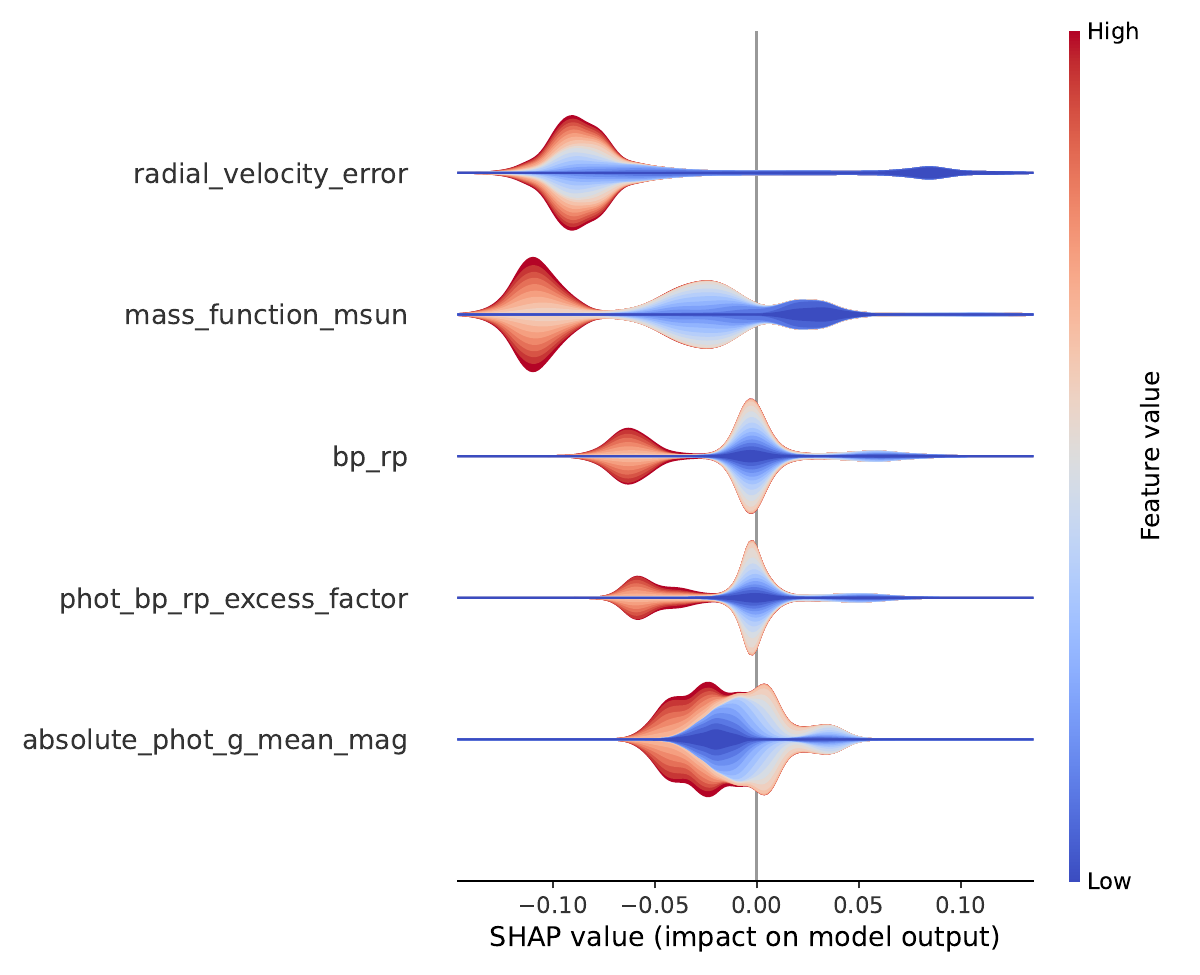}
\caption{SHAP value distribution for the top five features, {i.e.\ the input parameters such as radial velocity error}, in a random sample of 50 sources per configuration; colour indicates distribution of the feature; especially larger values of the variables are associated with smaller SHAP values. {The top five features are determined as having the highest average over the entire sample.} \label{fig:random_shap}}
\end{figure}

In terms of understanding the predictions on the entire population and the candidate pool, Figures \ref{fig:candidate_shap} and \ref{fig:random_shap} show the SHAP value distributions for the top five features of the candidates and a random sample of sources, respectively. The colour indicates the feature {value,  e.g.\ the radial velocity error. Notably, only smaller feature values of the mass function (mass\_function\_msun) and radial velocity error are strongly associated with higher SHAP values, indicating a greater influence on classifying a sample as an outlier. This trend suggests that certain features are particularly important when their values are small.}

\begin{figure} 
\centering
\includegraphics[width=\linewidth]{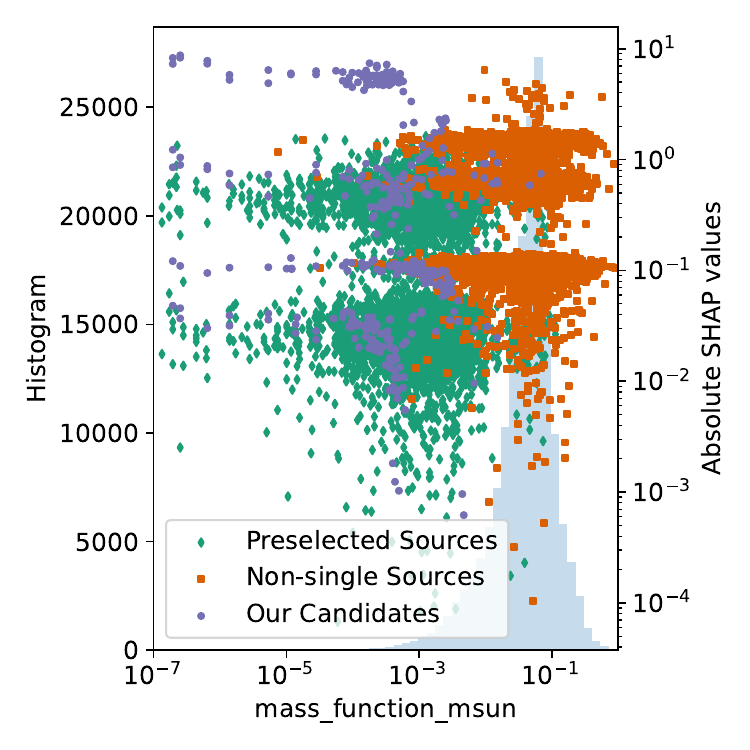}
\caption{SHAP values for mass\_function\_msun of top candidates identified by our approach vs. a random sample of 50 sources per configuration (total number of 8$\times$50 sources) in relation to their mass\_function\_msun values and in the background a histogram of the value distribution in the non-single source population \label{fig:mfun_shap}; Note that all groups show no trivial relationship between small mass\_function\_msun values and high SHAP values but form clusters}
\end{figure}

A detailed look at individual SHAP values in association with mass\_function\_msun values is available in Figure \ref{fig:mfun_shap} which highlights their complex interaction. 
Conversely, the random sample in Fig.\ \ref{fig:random_shap} displays that most samples are ruled out as candidates by large mass\_function\_msun values and/or large radial velocity errors. The \gaia colour bp\_rp of the source also seems to play a more important role here. Note, however the difference in magnitude on the x-axes of Figures \ref{fig:candidate_shap} and \ref{fig:random_shap}. 
Additionally, Figure \ref{fig:mfun_shap} also displays the distributions of SHAP values and mass\_function\_msun values in random samples of non-single source population and the \sscc. Note that even though they feature small mass\_function\_msun values, these are not associated with large SHAP values in most \sscc.

\subsubsection{Predictions for individual systems}
To investigate individual sources/solutions, we want to highlight three specific samples: The missed confirmed brown dwarf in LHS 1610, the top candidate around HD 40503, and \gaia DR3 246890014559489792 (\gaia object of interest identifier Gaia-ASOI-016), a small mass\_function\_msun outlier that was not identified as a candidate by our analysis. This can help us understand the decision process of the models, find potential confounders and help interpret the data of individual samples. In addition, decision plots for the missed brown dwarf around HD 89707 are in the Appendix in Figures \ref{fig:feat_bd2_rf} and \ref{fig:feat_bd2_xgb}. It was likely just barely below the detection threshold scoring highly on multiple models.

\begin{figure} 
\centering
\includegraphics[width=\linewidth]{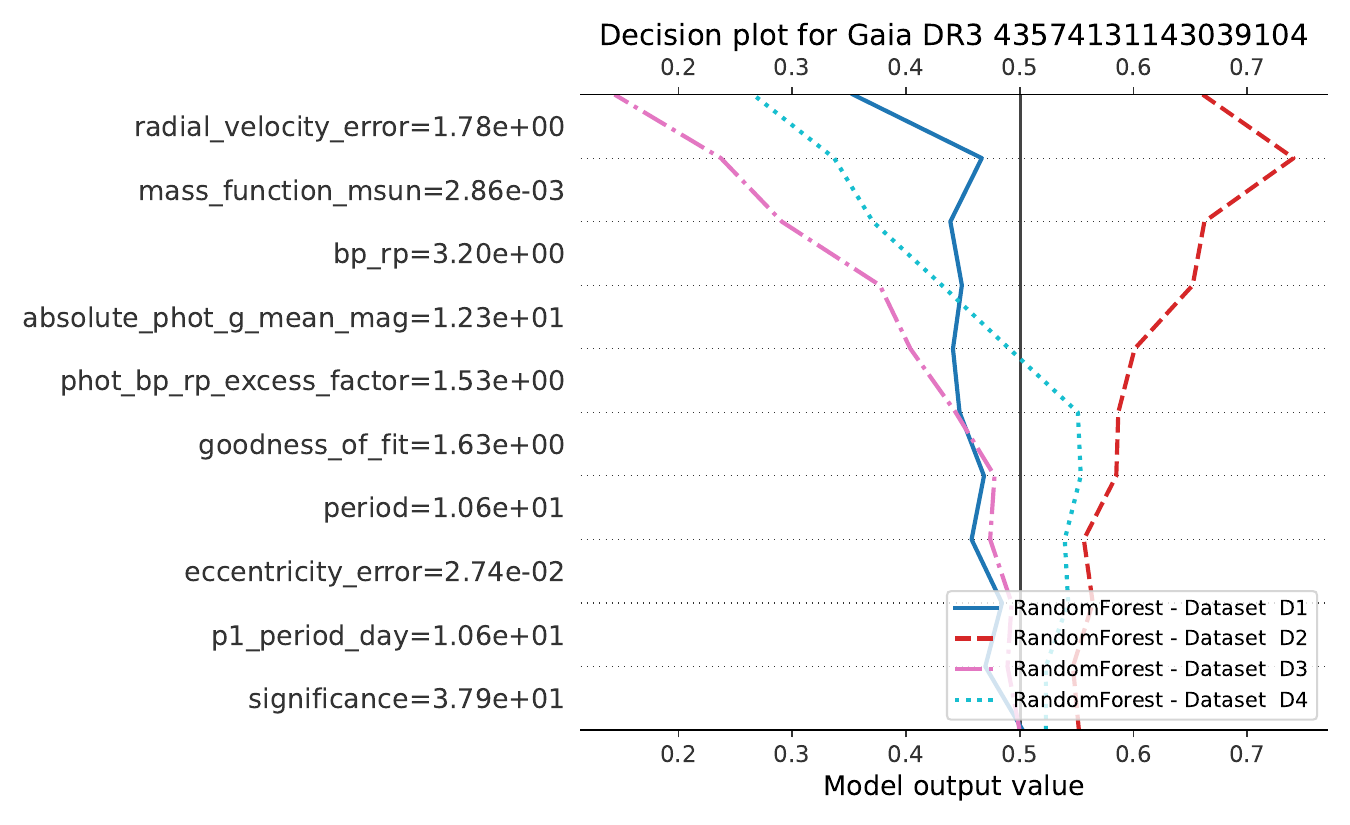}
\caption{Decision plot for the random forest models for the known but not identified brown dwarf in \gaia DR3 43574131143039104 / LHS 1610 \citep{2024AJ....168..140F}; high radial-velocity error, mass\_function\_msun and bp\_rp are the main contributors to the decision against an outliers classification for the models trained on D1, D3 and D4. \label{fig:feat_bd_rf}}
\end{figure}

\begin{figure} 
\centering
\includegraphics[width=\linewidth]{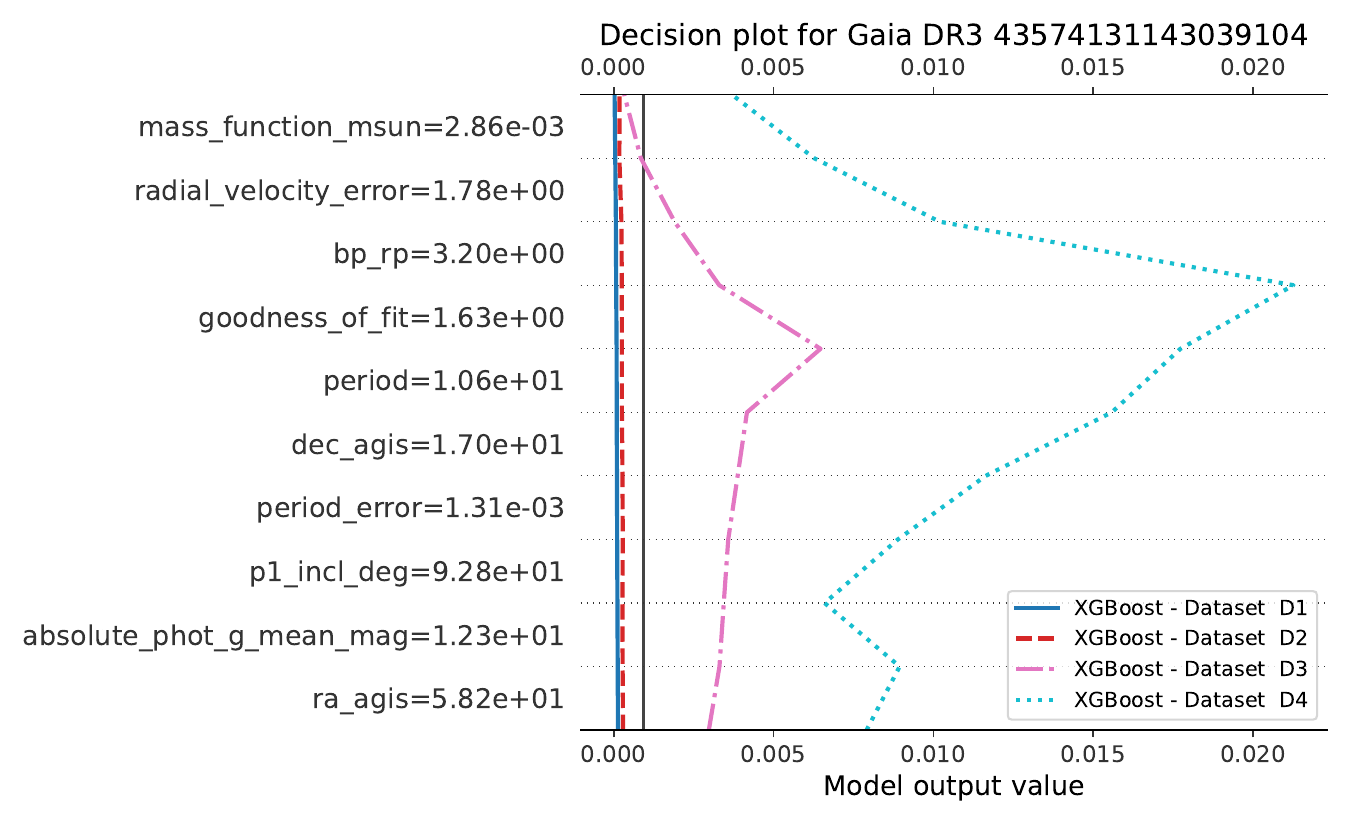}
\caption{Decision plot for the XGBoost models for the known but not identified brown dwarf in \gaia DR3 43574131143039104 / LHS 1610 (cf.\ Figure \ref{fig:feat_bd_rf}); high radial-velocity error, mass\_function\_msun and bp\_rp are the main contributors to the decision against an outliers classification for the models trained on D1, D3 and D4. \label{fig:feat_bd_xgb}}
\end{figure}

\begin{figure} 
\centering
\includegraphics[width=\linewidth]{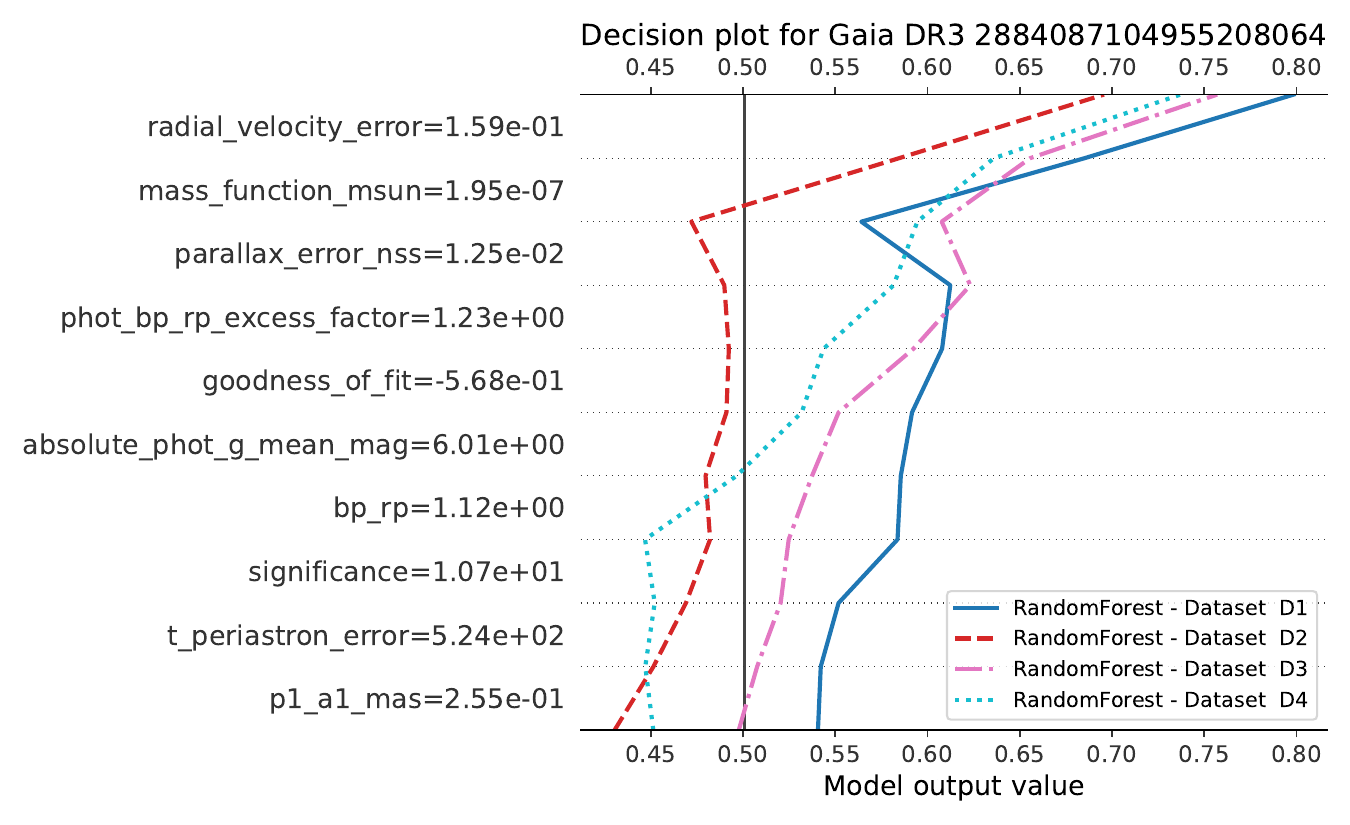}
\caption{Decision plot for the random forest models for top candidate \gaia DR3 2884087104955208064 / HD 40503. Small values of the radial-velocity error and mass\_function\_msun were the most decisive features. Notably, the bp\_rp colour was not an important feature. \label{fig:feat_top_rf}}
\end{figure}

\begin{figure} 
\centering
\includegraphics[width=\linewidth]{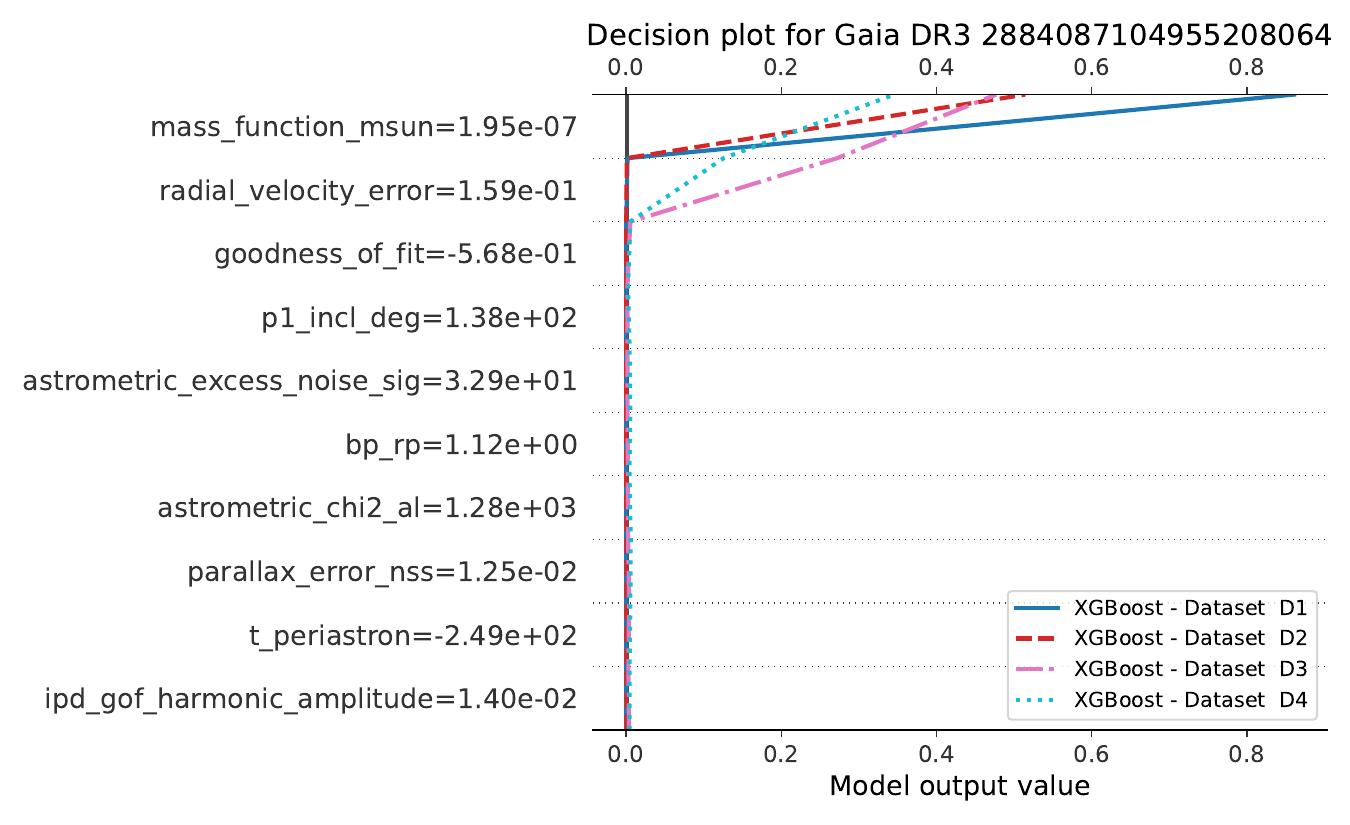}
\caption{Decision plot for the XGBoost models for top candidate \gaia DR3 2884087104955208064 / HD 40503. Small values of the radial-velocity error and mass\_function\_msun were the most decisive features. Notably, the bp\_rp colour was not an important feature. \label{fig:feat_top_xgb}}
\end{figure}

\begin{figure} 
\centering
\includegraphics[width=\linewidth]{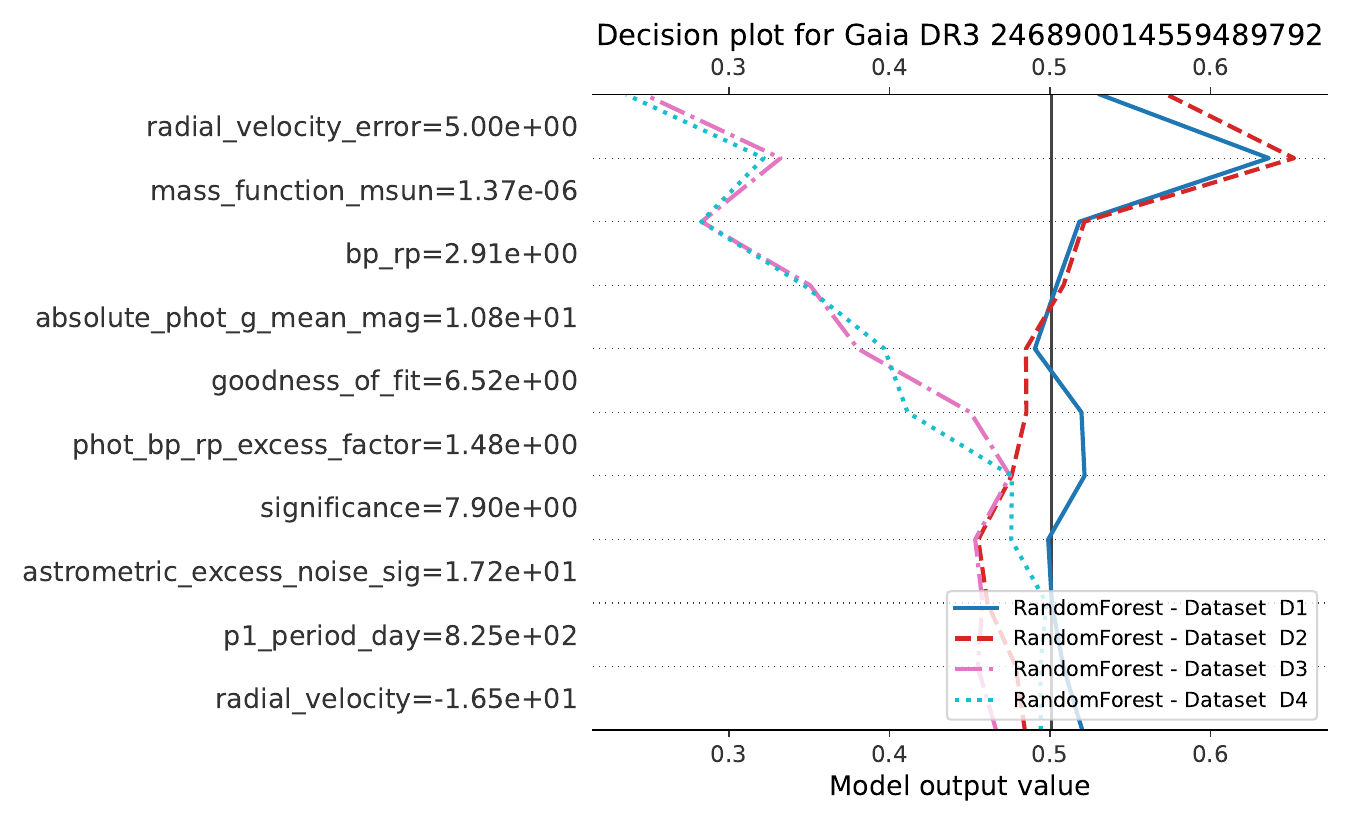}
\caption{Decision plot for the random forest models for \gaia DR3 246890014559489792 / Gaia-ASOI-016, a candidate with low mass\_function\_msun. Notably, the other features - especially the radial-velocity error and bp\_rp -  lead to an inlier classification. \label{fig:feat_lowmfun_rf}}
\end{figure}

\begin{figure} 
\centering
\includegraphics[width=\linewidth]{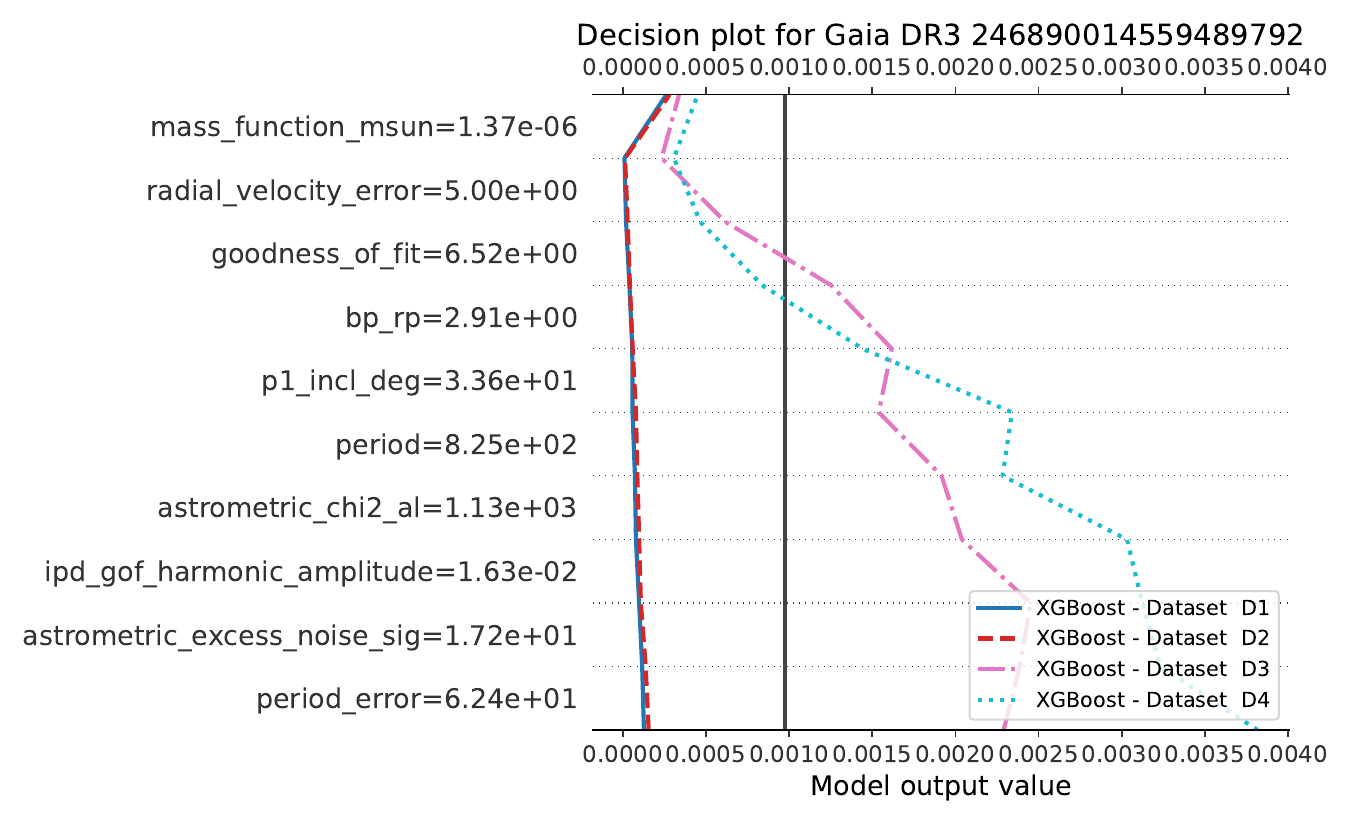}
\caption{Decision plot for the XGBoost models for \gaia DR3 246890014559489792 / Gaia-ASOI-016, a candidate with low mass\_function\_msun. Notably, the other features - especially the radial-velocity error, bp\_rp, and goodness-of-fit -  lead to an inlier classification.\label{fig:feat_lowmfun_xgb}}
\end{figure}

{Additionally, Figures \ref{fig:feat_bd_rf} - \ref{fig:feat_lowmfun_xgb} present decision plots that illustrate the contributions of individual features to the model's prediction. Each row corresponds to a specific feature, and the features are sorted by their overall importance averaged across models. The feature values are displayed on the left-hand side of the plot. For each feature, the coloured line shows how it contributes to the model's prediction. A positive slope, i.e.\ an upwards rightward displacement, of the line within a row indicates a positive contribution to the classification of the sample as an outlier like for, e.g., an exoplanet or brown dwarf. Conversely, a negative slope, i.e.\ and upwards leftward displacement, indicates a negative contribution to the classification. The grey line at the centre of the plot represents the base value of the model's prediction before considering the effects of individual features. This visualisation helps identifying which features drive the model's decision and how their specific values influence the output.}

Figures \ref{fig:feat_bd_rf} and \ref{fig:feat_bd_xgb} show decision plots for the missed brown dwarf around LHS 1610 for the random forest and XGBoost models, respectively. Here, the high radial-velocity error and mass\_function\_msun compared to other candidates such as HD\,40503 are the main contributors to the decision. However, we can also see that the colour (bp\_rp = 3.2) is a strong contributor, which makes sense as this is one of only two confirmed outliers with such a red colour - all others have bp\_rp $<$ 1.5 as seen in Figure \ref{fig:overview}. Overall, we can indeed see that with those characteristics the brown dwarf around LHS 1610 is atypical compared to the other confirmed brown dwarfs and exoplanets and the model's decision process is stringent. A remedy to this mis-identification would be the inclusion of more labelled training samples like the confirmed brown dwarf around LHS 1610.

Figures \ref{fig:feat_top_rf} and \ref{fig:feat_top_xgb} display the decision plots for the top exoplanet candidate around HD 40503 for the random forest and XGBoost models, respectively. The classification as outlier is driven by small values of the radial-velocity error and mass\_function\_msun. Notably, the bp\_rp colour was not an important feature.

Finally, Figures \ref{fig:feat_lowmfun_rf} and \ref{fig:feat_lowmfun_xgb} show the decision plots for \gaia DR3 246890014559489792 / Gaia-ASOI-016. Looking at the small mass\_function\_msun value of 1.367e-6 (cf. Figure \ref{fig:overview}) it would be conceivable that this is a strong candidate. However, as seen in the decision plots this is not the case because the other features, in particular the radial-velocity error, bp\_rp, and goodness-of-fit contribute to an inlier classification.

Additional data and a full list of candidates are given in the Appendix.

\section{Discussion}
The results shown in this work clearly indicate the viability of using a machine learning-based approach to identifying likely candidates for exoplanets and brown-dwarf companions. The majority of known exoplanets and brown-dwarf companions in the dataset were picked up successfully by the models and the identified new candidates exhibit physical properties, such as expected companion mass, that indicate that they indeed are worthy of follow-up studies.  

One predictable drawback of the approach in this regard is the tendency of identifying new candidates that mostly lie within the distribution of already-known candidates in terms of their features. For example, the location of our best candidates in the colour-magnitude diagram (Fig.\ \ref{fig:overview} top) largely coincides with the region where the confirmed substellar companions, i.e.\ the training sample, resides. This is a common drawback of machine learning approaches and could be remedied by providing a richer `training` dataset that, e.g., includes more confirmed substellar companions around redder sources (bp\_rp > 1.5). In the same vein it has to be mentioned that the presented approach is of course greatly impacted by the very limited number of confirmed samples. These aspects will improve with time as the results of various \gaia DR3-orbit follow-up studies are being published and certainly with the release of \gaia DR4, when many more orbits of stars with substellar companions will be released. We anticipate that the iterative identification and confirmation of an increasing number of exoplanets with \gaia orbits, covering growing areas in parameter space, will lead to a positive feedback loop with machine learning approaches, which will help to accelerate the pace of discoveries.

A particularly precious dataset for approaches like ours are small mass-function systems that actually are confirmed to be near-twin binaries \citep{2023AJ....165..266M}. These systems are crucial for \emph{teaching} the models how to identify these astrophysical false positives. We therefore urge the teams involved in the observational follow-up of \gaia astrometric exoplanet candidates to publish their results, regardless of the outcome.

In terms of the performance and power of the presented approach several observations can be made. First, as displayed in Figures \ref{fig:candidate_shap} and \ref{fig:random_shap}, the decision making of the models is truly multivariate as different features can lead to the identification of likely outliers and different features can support either a classification as outlier or not. For example, the goodness-of-fit feature is the third strongest in identification of our candidates but does not appear among the random sample were instead bp\_rp is third strongest. 

Another indicator of such behaviour can be seen in Figure \ref{fig:mfun_shap}. Even though mass\_function\_msun is the strongest feature overall (with small values being associated with a classification as outlier) a small value alone is not sufficient for a classification as outlier as exemplified by \gaia DR3 246890014559489792 / Gaia-ASOI-016 and in Figures \ref{fig:feat_lowmfun_rf} and \ref{fig:feat_lowmfun_xgb}. 

These kinds of observations are interesting for complementing a human-based analysis of the data since one can use the models' identification of important features to identify and investigate potential confounders and spurious relationships in the data. In some cases, such as for the eliminated features like parallax and absolute source magnitude, this is trivial but in others it can inspire follow-up investigations. 

\begin{figure} 
\centering
\includegraphics[width=\linewidth]{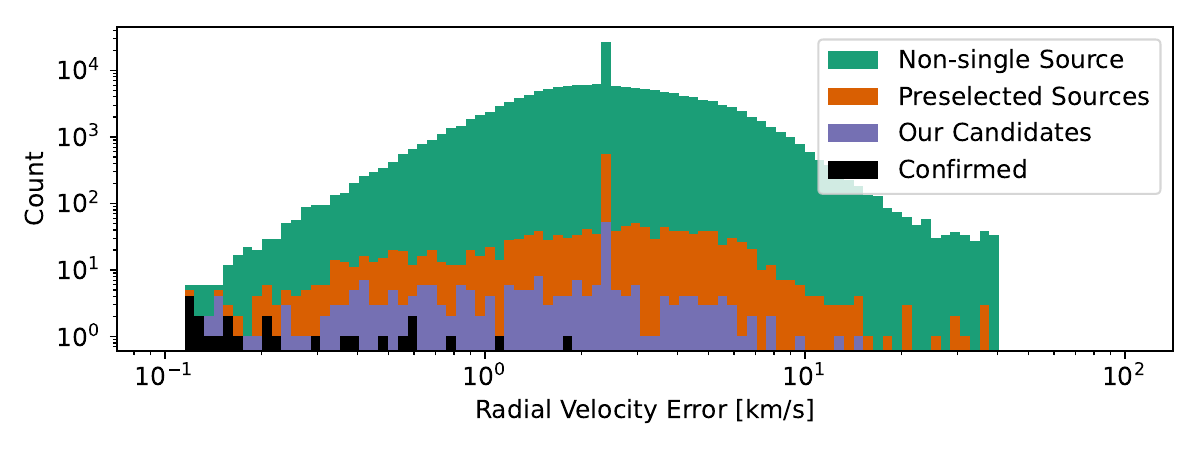}
\caption{Distribution of radial velocity error in the sub-populations. The central peaks are caused by data imputation (Section \ref{sec:imputation}) with the median for missing RV error values.}
\label{fig:rv_hist}
\end{figure}

The most obvious example of this is the radial\_velocity\_error. The models consistently identified it as a strong feature and, as displayed in Figure \ref{fig:rv_hist}, there is a correlation between the presence of a confirmed exoplanet or brown-dwarf companion and a smaller radial\_velocity\_error. As explained in Section \ref{sec:selected_features}, this particular feature has a complex (expected) relation with the presence of orbital signatures. All hosts of confirmed substellar companions have RV errors smaller than the median value for astrometric orbits (Fig. \ref{fig:rv_hist}). The same applies to almost all our identified candidates. The underlying reason is expected to be a mixture of outlier classification (for small RV errors) and inlier classification (for large RV errors that can be caused by unmodelled binary motion). A detailed investigation of these effects is outside the scope of this initial proof of concept. We note, however, that we performed ablation studies that indicated that our approach identifies similar candidates without the  radial\_velocity\_error feature. Thus, even if this relationship is spurious, it should not compromise the performance of the presented approach.

\section{Conclusions}
We successfully demonstrated a new approach for identifying individual systems with particular properties, i.e.\ outliers, in the ensemble of \gaia DR3 astrometric orbits. As first application, we searched for the most likely sources orbited by extrasolar planets and brown dwarfs, but the approach could be adapted to other science cases, e.g.\ to search for black holes or twin binary stars instead.

Using machine learning, we identified 22 high-confidence systems with substellar companions, of which five host extrasolar-planet candidates that were previously known. Two of those candidates were invalidated during the review process of this paper because the corresponding \gaia orbital solutions were retracted as false positives. We identified one of the exoplanet candidates (BD+24 4592, Gaia-ASOI-005) as a false-positive because the \gaia orbit does not correspond to the true orbit of the system, which is a known RV binary. 

We identified 17 high-confidence systems with brown-dwarf-companion candidates. Two of those candidates (HD\,104289 and HD\,153376) are false positives because the \gaia orbits do not correspond to the true orbits of the known RV binaries, and one candidate (BD-06 2423A) has in fact a low-mass stellar companion although it was picked up by our models. Three brown-dwarf companion candidates (HD 206484, HD 78631, GSC 09436-0108) have combined astrometric+RV orbital solutions (\typeASB) and the companions straddle the substellar-stellar mass boundary. The companion of HD\,106888 is a very low-mass star. 

The presence of a brown dwarf with an astrometric mass of ${M_2} = {62.43}^{+6.16}_{-5.44}\,{M_\mathrm{Jup}}$ orbiting our high-confidence candidate G 15-6 was confirmed by literature RV measurements. We propose that the remaining nine high-confidence brown-dwarf companion candidates identified in this study be followed-up preferentially, e.g.\ with radial velocities or interferometric imaging \citep[e.g.][]{2024A&A...688A..44W}.

As the \gaia astrometric-orbit catalogue will grow significantly in DR4, machine learning approaches will become more and more important for prioritising follow-up and identifying interesting systems and `outliers`. In the same way, any analysis of the DR4 epoch data (\gaia astrometric timeseries of all sources will be released with DR4\footnote{\url{https://www.cosmos.esa.int/web/gaia/release}}) that has the goal of identifying non-single stars, e.g.\ through non-linear motions and drifts, single-, or multi-companion orbits, will have to address the problem of false-positives. The holistic machine learning approach that facilitates the use of as much information as available has many advantages in that respect.
In particular, once sufficient training data will be available, it is very likely that machine learning algorithms will automatically identify a sizeable fraction of twin binary stars, which are the principal astrophysical false positives that affect \gaia astrometric planet searches.

As a byproduct of this study and as part of the validation of our best candidates we uncovered that the astrometric orbits for 3 out of 22 solutions ($\sim$$14^{+10}_{-5}$ \% using binomial statistics) are false positives. They occurred when the \gaia DR3 astrometric exoplanet pipeline \citep{2023AA...674A..10H}\footnote{This pipeline produced only solutions of type \typeOrbTarStar and \typeOrbAltStar.} processed the \gaia data collected over $\sim$1000 days for sources that are large-amplitude binary stars with much-longer true periods between 2400 and 4900 days.
The incomplete orbital arc was then fitted with a small-amplitude signal with a period shorter than the true period, which resulted in a candidate system with substellar companion. 
Whereas the tendency of astrometric pipelines to fit incompletely-covered orbits with short-period spurious orbital solutions has been discussed previously \citep{Casertano:2008th,2023AA...674A..10H}, the particularly damaging effect this may have on small-amplitude solutions that can be mistaken for the signatures of exoplanets has to our knowledge not been documented before. This issue is likely to be partially mitigated in \gaia DR4 with its superior data quality, quantity, and time span.

\section*{Acknowledgements}
This work has made use of data from the European Space Agency (ESA) mission {\it Gaia} (\url{https://www.cosmos.esa.int/gaia}), processed by the {\it Gaia} Data Processing and Analysis Consortium (DPAC, \url{https://www.cosmos.esa.int/web/gaia/dpac/consortium}). Funding for the DPAC has been provided by national institutions, in particular the institutions participating in the {\it Gaia} Multilateral Agreement.

This research has made use of the SIMBAD database, operated at CDS, Strasbourg, France \citep{Wenger:2000ve}. The authors made use of \texttt{astropy} (a community-developed core Python package for Astronomy \citep{Astropy-Collaboration:2013aa}, \texttt{scipy} \citep{Jones:2001aa}, \texttt{numpy} \citep{Oliphant2007}, \texttt{ipython} \citep{Perez2007}, \texttt{pandas} \citep{jeff_reback_2022_6408044}, \texttt{matplotlib} \citep{hunter2007}, and \texttt{astroquery} \citep{2019AJ....157...98G}
This research made use of \texttt{pystrometry}, an open source Python package for astrometry timeseries analysis \citep{johannes_sahlmann_2019_3515526}.
This project made use of the ESA Datalabs platform.

The authors acknowledge the positive role of the ESAC canteen and its friendly staff for facilitating informal exchanges during breakfast. This project is the outcome of one of these.

%%%%%%%%%%%%%%%%%%%%%%%%%%%%%%%%%%%%%%%%%%%%%%%%%%
\section*{Data Availability}
The \gaia data underlying this research are available at the Gaia archive at ESA (DOI:  \url{https://doi.org/10.5270/esa-qa4lep3}). The relevant tables are \texttt{gaiadr3.}\gstable, \texttt{gaiadr3.}\nssTBO, and \texttt{gaiadr3.}\nssBM.

The code and data needed to reproduce our results are available at \url{https://github.com/esa/gaia-astrometric-exoplanet-orbit-ml}. 

% The inclusion of a Data Availability Statement is a requirement for articles published in MNRAS. Data Availability Statements provide a standardised format for readers to understand the availability of data underlying the research results described in the article. The statement may refer to original data generated in the course of the study or to third-party data analysed in the article. The statement should describe and provide means of access, where possible, by linking to the data or providing the required accession numbers for the relevant databases or DOIs.



\bibliographystyle{mnras}
\bibliography{biblio.bib} 

%%%%%%%%%%%%%%%%%%%%%%%%%%%%%%%%%%%%%%%%%%%%%%%%%%

%%%%%%%%%%%%%%%%% APPENDICES %%%%%%%%%%%%%%%%%%%%%

\appendix
\section{Auxiliary tables}
\begin{table}
\centering
\caption{Used and computed fields for the \gstable table. Instead of reproducing the field descriptions from \citet{2022gdr3.reptE....V} here, we included direct links to the online documentation. Columns commented as `not used` and `confounder` were not used in the analyses. The latter correspond to features that are marked as important for identifying outliers but that are not expected as such based on physical arguments, see Section \ref{sec:features}.} \label{tab:gs_columns}
\begin{tabular}{rp{3.0cm}r}
\hline
Column & Description & Comment \\
\hline
source\_id & \href{https://gea.esac.esa.int/archive/documentation/GDR3/Gaia_archive/chap_datamodel/sec_dm_main_source_catalogue/ssec_dm_gaia_source.html#gaia_source-source_id}{Identifier} & not used \\
ra & \href{https://gea.esac.esa.int/archive/documentation/GDR3/Gaia_archive/chap_datamodel/sec_dm_main_source_catalogue/ssec_dm_gaia_source.html#gaia_source-ra}{Right Ascension (RA)} &  \\
dec & \href{https://gea.esac.esa.int/archive/documentation/GDR3/Gaia_archive/chap_datamodel/sec_dm_main_source_catalogue/ssec_dm_gaia_source.html#gaia_source-dec}{Declination (Dec)} &  \\
parallax & \href{https://gea.esac.esa.int/archive/documentation/GDR3/Gaia_archive/chap_datamodel/sec_dm_main_source_catalogue/ssec_dm_gaia_source.html#gaia_source-parallax}{Trigonometric parallax} & confounder \\
parallax\_error & \href{https://gea.esac.esa.int/archive/documentation/GDR3/Gaia_archive/chap_datamodel/sec_dm_main_source_catalogue/ssec_dm_gaia_source.html#gaia_source-parallax_error}{Standard error} &  \\
pmra & \href{https://gea.esac.esa.int/archive/documentation/GDR3/Gaia_archive/chap_datamodel/sec_dm_main_source_catalogue/ssec_dm_gaia_source.html#gaia_source-pmra}{Proper motion in RA} & confounder \\
pmdec & \href{https://gea.esac.esa.int/archive/documentation/GDR3/Gaia_archive/chap_datamodel/sec_dm_main_source_catalogue/ssec_dm_gaia_source.html#gaia_source-pmdec}{Proper motion in Dec} & confounder \\
visibility\_periods\_used & \href{https://gea.esac.esa.int/archive/documentation/GDR3/Gaia_archive/chap_datamodel/sec_dm_main_source_catalogue/ssec_dm_gaia_source.html#gaia_source-visibility_periods_used}{Number of independent measurements} &  \\
radial\_velocity & \href{https://gea.esac.esa.int/archive/documentation/GDR3/Gaia_archive/chap_datamodel/sec_dm_main_source_catalogue/ssec_dm_gaia_source.html#gaia_source-radial_velocity}{Radial velocity} &  \\
radial\_velocity\_error & \href{https://gea.esac.esa.int/archive/documentation/GDR3/Gaia_archive/chap_datamodel/sec_dm_main_source_catalogue/ssec_dm_gaia_source.html#gaia_source-radial_velocity_error}{Standard error} &  \\
astrometric\_chi2\_al & \href{https://gea.esac.esa.int/archive/documentation/GDR3/Gaia_archive/chap_datamodel/sec_dm_main_source_catalogue/ssec_dm_gaia_source.html#gaia_source-astrometric_chi2_al}{Fit-quality metric} &  \\
ipd\_gof\_harmonic\_amplitude & \href{https://gea.esac.esa.int/archive/documentation/GDR3/Gaia_archive/chap_datamodel/sec_dm_main_source_catalogue/ssec_dm_gaia_source.html#gaia_source-ipd_gof_harmonic_amplitude}{Statistic related to scan-angle dependence} &  \\
bp\_rp & \href{https://gea.esac.esa.int/archive/documentation/GDR3/Gaia_archive/chap_datamodel/sec_dm_main_source_catalogue/ssec_dm_gaia_source.html#gaia_source-bp_rp}{Stellar colour} &  \\
phot\_bp\_rp\_excess\_factor & \href{https://gea.esac.esa.int/archive/documentation/GDR3/Gaia_archive/chap_datamodel/sec_dm_main_source_catalogue/ssec_dm_gaia_source.html#gaia_source-phot_bp_rp_excess_factor}{Colour excess factor} &  \\
ruwe & \href{https://gea.esac.esa.int/archive/documentation/GDR3/Gaia_archive/chap_datamodel/sec_dm_main_source_catalogue/ssec_dm_gaia_source.html#gaia_source-ruwe}{Renormalised fit-quality} &  \\
phot\_g\_mean\_mag & \href{https://gea.esac.esa.int/archive/documentation/GDR3/Gaia_archive/chap_datamodel/sec_dm_main_source_catalogue/ssec_dm_gaia_source.html#gaia_source-phot_g_mean_mag}{Apparent magnitude in $G$} & confounder \\
phot\_rp\_mean\_mag & \href{https://gea.esac.esa.int/archive/documentation/GDR3/Gaia_archive/chap_datamodel/sec_dm_main_source_catalogue/ssec_dm_gaia_source.html#gaia_source-phot_rp_mean_mag}{Red \gaia magnitude ()} &  \\
phot\_bp\_mean\_mag & \href{https://gea.esac.esa.int/archive/documentation/GDR3/Gaia_archive/chap_datamodel/sec_dm_main_source_catalogue/ssec_dm_gaia_source.html#gaia_source-phot_bp_mean_mag}{Blue \gaia magnitude ()} &  \\
astrometric\_excess\_noise & \href{https://gea.esac.esa.int/archive/documentation/GDR3/Gaia_archive/chap_datamodel/sec_dm_main_source_catalogue/ssec_dm_gaia_source.html#gaia_source-astrometric_excess_noise}{Source-specific excess noise} &  \\
astrometric\_excess\_noise\_sig & \href{https://gea.esac.esa.int/archive/documentation/GDR3/Gaia_archive/chap_datamodel/sec_dm_main_source_catalogue/ssec_dm_gaia_source.html#gaia_source-astrometric_excess_noise_sig}{Excess noise significance} &  \\
astrometric\_primary\_flag & \href{https://gea.esac.esa.int/archive/documentation/GDR3/Gaia_archive/chap_datamodel/sec_dm_main_source_catalogue/ssec_dm_gaia_source.html#gaia_source-astrometric_primary_flag}{Primary source flag} & not used \\
astrometric\_n\_good\_obs\_al & \href{https://gea.esac.esa.int/archive/documentation/GDR3/Gaia_archive/chap_datamodel/sec_dm_main_source_catalogue/ssec_dm_gaia_source.html#gaia_source-astrometric_n_good_obs_al}{Number of used measurements} &  \\
absolute\_phot\_g\_mean\_mag & absolute $G$ mag & computed \\
\hline
\end{tabular}
\end{table}
\begin{table}
\centering
\caption{Used and computed fields for the \nssTBO table.} \label{tab:tbo_columns}
\begin{tabular}{rp{2.8cm}r}
\hline
Column & Description & Comment \\
\hline
source\_id & \href{https://gea.esac.esa.int/archive/documentation/GDR3/Gaia_archive/chap_datamodel/sec_dm_non--single_stars_tables/ssec_dm_nss_two_body_orbit.html#nss_two_body_orbit-source_id}{Identifier} & not used \\
nss\_solution\_type & \href{https://gea.esac.esa.int/archive/documentation/GDR3/Gaia_archive/chap_datamodel/sec_dm_non--single_stars_tables/ssec_dm_nss_two_body_orbit.html#nss_two_body_orbit-nss_solution_type}{Solution type} & not used \\
ra & \href{https://gea.esac.esa.int/archive/documentation/GDR3/Gaia_archive/chap_datamodel/sec_dm_non--single_stars_tables/ssec_dm_nss_two_body_orbit.html#nss_two_body_orbit-ra}{RA} &  \\
dec & \href{https://gea.esac.esa.int/archive/documentation/GDR3/Gaia_archive/chap_datamodel/sec_dm_non--single_stars_tables/ssec_dm_nss_two_body_orbit.html#nss_two_body_orbit-dec}{Dec} &  \\
parallax & \href{https://gea.esac.esa.int/archive/documentation/GDR3/Gaia_archive/chap_datamodel/sec_dm_non--single_stars_tables/ssec_dm_nss_two_body_orbit.html#nss_two_body_orbit-parallax}{Trigonometric parallax} & confounder \\
parallax\_error & \href{https://gea.esac.esa.int/archive/documentation/GDR3/Gaia_archive/chap_datamodel/sec_dm_non--single_stars_tables/ssec_dm_nss_two_body_orbit.html#nss_two_body_orbit-parallax_error}{Standard error} &  \\
pmra & \href{https://gea.esac.esa.int/archive/documentation/GDR3/Gaia_archive/chap_datamodel/sec_dm_non--single_stars_tables/ssec_dm_nss_two_body_orbit.html#nss_two_body_orbit-pmra}{Proper motion in RA} & confounder \\
pmdec & \href{https://gea.esac.esa.int/archive/documentation/GDR3/Gaia_archive/chap_datamodel/sec_dm_non--single_stars_tables/ssec_dm_nss_two_body_orbit.html#nss_two_body_orbit-pmdec}{Proper motion in Dec} & confounder \\
period & \href{https://gea.esac.esa.int/archive/documentation/GDR3/Gaia_archive/chap_datamodel/sec_dm_non--single_stars_tables/ssec_dm_nss_two_body_orbit.html#nss_two_body_orbit-period}{Orbital period} &  \\
period\_error & \href{https://gea.esac.esa.int/archive/documentation/GDR3/Gaia_archive/chap_datamodel/sec_dm_non--single_stars_tables/ssec_dm_nss_two_body_orbit.html#nss_two_body_orbit-period_error}{Standard error} &  \\
t\_periastron & \href{https://gea.esac.esa.int/archive/documentation/GDR3/Gaia_archive/chap_datamodel/sec_dm_non--single_stars_tables/ssec_dm_nss_two_body_orbit.html#nss_two_body_orbit-t_periastron}{Periastron time} &  \\
t\_periastron\_error & \href{https://gea.esac.esa.int/archive/documentation/GDR3/Gaia_archive/chap_datamodel/sec_dm_non--single_stars_tables/ssec_dm_nss_two_body_orbit.html#nss_two_body_orbit-t_periastron_error}{Standard error} &  \\
eccentricity & \href{https://gea.esac.esa.int/archive/documentation/GDR3/Gaia_archive/chap_datamodel/sec_dm_non--single_stars_tables/ssec_dm_nss_two_body_orbit.html#nss_two_body_orbit-eccentricity}{Orbit eccentricity} &  \\
eccentricity\_error & \href{https://gea.esac.esa.int/archive/documentation/GDR3/Gaia_archive/chap_datamodel/sec_dm_non--single_stars_tables/ssec_dm_nss_two_body_orbit.html#nss_two_body_orbit-eccentricity_error}{Standard error} &  \\
astrometric\_n\_good\_obs\_al & \href{https://gea.esac.esa.int/archive/documentation/GDR3/Gaia_archive/chap_datamodel/sec_dm_non--single_stars_tables/ssec_dm_nss_two_body_orbit.html#nss_two_body_orbit-astrometric_n_good_obs_al}{Number of used measurements} &  \\
obj\_func & \href{https://gea.esac.esa.int/archive/documentation/GDR3/Gaia_archive/chap_datamodel/sec_dm_non--single_stars_tables/ssec_dm_nss_two_body_orbit.html#nss_two_body_orbit-obj_func}{$\chi^2$ metric} &  \\
goodness\_of\_fit & \href{https://gea.esac.esa.int/archive/documentation/GDR3/Gaia_archive/chap_datamodel/sec_dm_non--single_stars_tables/ssec_dm_nss_two_body_orbit.html#nss_two_body_orbit-goodness_of_fit}{Gaussianised $\chi^2$ metric} &  \\
efficiency & \href{https://gea.esac.esa.int/archive/documentation/GDR3/Gaia_archive/chap_datamodel/sec_dm_non--single_stars_tables/ssec_dm_nss_two_body_orbit.html#nss_two_body_orbit-efficiency}{Correlation metric} &  \\
significance & \href{https://gea.esac.esa.int/archive/documentation/GDR3/Gaia_archive/chap_datamodel/sec_dm_non--single_stars_tables/ssec_dm_nss_two_body_orbit.html#nss_two_body_orbit-significance}{Orbit significance $=a_0/\sigma_{a_0}$} &  \\
p1\_a0\_mas & $a_0$ (mas) & computed \\
p1\_omega\_deg & $\omega$ (deg) & computed \\
p1\_OMEGA\_deg & $\Omega$ (deg) & computed \\
p1\_incl\_deg & $i$ (deg) & computed \\
mass\_function\_msun & $f_M$ ($M_{\sun}$) & computed \\
\hline
\end{tabular}
\end{table}
\begin{table*}
\centering
\caption{Assigned labels}
\label{tab:labeldef}
\footnotesize    
\begin{tabular}{rrrr}
\hline
Gaia DR3 source\_id & Name & label & Reference \\
\hline
2047188847334279424 & HD185501 & \bs & \cite{2020AJ....159..233H} \\
5122670101678217728 & HD12357 & \bs & \cite{2023AJ....165..266M} \\
2052469973468984192 & Ross 1063 & \bs & \cite{2023AJ....165..266M} \\
685029558383335168 & HD77065 & \bdc & \cite{2023AA...674A..34G} \\
3751763647996317056 & HD89707 & \bdc & \cite{2023AA...674A..34G} \\
3750881083756656128 & HD91669 & \bdc & \cite{2023AA...674A..34G} \\
3309006602007842048 & HD30246 & \bdc & \cite{2023AA...674A..34G} \\
2778298280881817984 & HD5433   & \bdc & \cite{2023AA...674A..34G} \\
2651390587219807744 & BD-004475 & \bdc & \cite{2023AA...674A..34G} \\
5563001178343925376 & HD52756 & \bdc & \cite{2023AA...674A..34G} \\
824461960796102528 & HD82460 & \bdc & \cite{2023AA...674A..34G} \\
43574131143039104 & LHS1610 & \bdc & \cite{2024AJ....168..140F} \\
1035000055055287680 & HD68638A & \bdc & \cite{2023AA...680A..16U} \\
5999024986946599808 & CD-4610046 & \bdc & \cite{2023AA...680A..16U} \\
1181993180456516864 & HD132032 & \bdc & \cite{2023AA...680A..16U} \\
873616860770228352 & BD+291539  & \bdc & \cite{2023AA...674A..34G} \\
855523714036230016 & HD92320 & \bdc & \cite{2023AA...674A..34G} \\
6421118739093252224 & HD175167 & \exoplanet & \cite{2023AA...674A..34G} \\
3424193536079703808 & HD39392 & \exoplanet & \cite{2023MNRAS.526.5155S} \\
2603090003484152064 & GJ876 & \exoplanet & \cite{2023AA...674A..34G} \\
4976894960284258048 & HD142 & \exoplanet & \cite{2023AA...674A..34G} \\
637329067477530368 & HD81040 & \exoplanet & \cite{2023AA...674A..34G} \\
5855730584310531200 & HD111232 & \exoplanet & \cite{2023AA...674A..34G} \\
2367734656180397952 & BD-170063  & \exoplanet & \cite{2023AA...674A..34G} \\
4745373133284418816 & HR810 & \exoplanet & \cite{2023AA...674A..34G} \\
1594127865540229888 & HD132406 & \exoplanet & \cite{2023AA...674A..34G} \\
4062446910648807168 & HD164604 & \exoplanet & \cite{2023AA...674A..34G} \\
4698424845771339520 & WD 0141-675 & \fpo & Cosmos Pages \\
5765846127180770432 & HIP 64690 & \fpo & Cosmos Pages \\
5266148569447305600 & HD42936 & \vlmsc & \cite{2020NatAs...4..419B} \\
1224551770875466496 & HD140913 & \vlmsc & \cite{2023AA...680A..16U} \\
4753355209745022208 & HD17155 & \vlmsc & \cite{2023AA...680A..16U} \\
\hline
\end{tabular}
\end{table*}

\begin{table*}
\centering
\caption{All 66 lower-confidence candidates selected as having $0.125<\rho <= 0.5$. Colums are the same as in Table \ref{tab:candidates1}.}
\label{tab:candidates2}
\footnotesize    
\begin{tabular}{rrrrrrrrrrrrr}
\hline
\gaia DR3 source\_id & Name & SpT & sol.  & $P$ & $f_M$ & $M_2$ & $\rho$ & $\rho_\mathrm{ssc}$ & $\rho_\mathrm{nss}$ & ssc & $M_1$ & $M_{2, \mathrm{alt}}$\\
& & &type &(day) &($M_{\sun}$) & ($M_\mathrm{Jup}$) & & & & & ($M_{\sun}$) & ($M_{\sun}$)\\
\hline
932447162423519232 & \href{http://simbad.u-strasbg.fr/simbad/sim-basic?Ident=Gaia+DR3+932447162423519232&submit=SIMBAD+search}{UCAC4 684-047911} & M2.3 & ASB1 & 407.6 & 1.2e-05 & 18.0 & 0.2 & 0.0 & 0.5 & 0 & ${0.63}^{+0.05}_{-0.15}$ & $0.150$-$0.639$ \\
4632430611684633344 & \href{http://simbad.u-strasbg.fr/simbad/sim-basic?Ident=Gaia+DR3+4632430611684633344&submit=SIMBAD+search}{HD  14717} & G0V & Orbital & 778.4 & 1.4e-05 & 27.2 & 0.2 & 0.5 & 0.0 & 1 & ${1.10}^{+0.05}_{-0.10}$ & $0.022$-$1.208$ \\
4652393894560537728 & \href{http://simbad.u-strasbg.fr/simbad/sim-basic?Ident=Gaia+DR3+4652393894560537728&submit=SIMBAD+search}{HD  31251} & G2V & ASB1 & 419.8 & 8.1e-05 & 46.6 & 0.2 & 0.0 & 0.5 & 0 & NaN & NaN \\
2098419251579450880 & \href{http://simbad.u-strasbg.fr/simbad/sim-basic?Ident=Gaia+DR3+2098419251579450880&submit=SIMBAD+search}{2MASS J18405933+3950016} & -- & Orbital & 241.7 & 8.9e-05 & 36.3 & 0.2 & 0.2 & 0.2 & 1 & ${0.65}^{+0.16}_{-0.16}$ & $0.029$-$0.936$ \\
2984269003840680832 & \href{http://simbad.u-strasbg.fr/simbad/sim-basic?Ident=Gaia+DR3+2984269003840680832&submit=SIMBAD+search}{BD-15  1060} & G5 & Orbital & 424.9 & 9.5e-05 & 52.2 & 0.2 & 0.5 & 0.0 & 1 & ${1.09}^{+0.06}_{-0.09}$ & $0.045$-$1.291$ \\
3205578257602278784 & \href{http://simbad.u-strasbg.fr/simbad/sim-basic?Ident=Gaia+DR3+3205578257602278784&submit=SIMBAD+search}{TYC 4730-512-1} & -- & Orbital & 598.1 & 9.7e-05 & 49.1 & 0.2 & 0.2 & 0.2 & 1 & ${0.98}^{+0.06}_{-0.09}$ & $0.041$-$1.171$ \\
2994437527894182272 & \href{http://simbad.u-strasbg.fr/simbad/sim-basic?Ident=Gaia+DR3+2994437527894182272&submit=SIMBAD+search}{HD  42606} & K3V & OTS & 800.1 & 9.8e-05 & 46.4 & 0.2 & 0.0 & 0.5 & 0 & ${0.90}^{+0.05}_{-0.16}$ & $0.516$-$0.984$ \\
5981932494567274368 & \href{http://simbad.u-strasbg.fr/simbad/sim-basic?Ident=Gaia+DR3+5981932494567274368&submit=SIMBAD+search}{CD-51  9508} & G8 & ASB1 & 413.9 & 1.0e-04 & 48.7 & 0.2 & 0.5 & 0.0 & 1 & ${0.96}^{+0.06}_{-0.06}$ & ${0.044}^{+0.008}_{-0.008}$ \\
4854976609869179648 & \href{http://simbad.u-strasbg.fr/simbad/sim-basic?Ident=Gaia+DR3+4854976609869179648&submit=SIMBAD+search}{CD-41  1115} & G5 & ASB1 & 406.5 & 1.2e-04 & 50.6 & 0.2 & 0.5 & 0.0 & 1 & ${0.94}^{+0.06}_{-0.06}$ & ${0.068}^{+0.008}_{-0.009}$ \\
2217319511295868672 & \href{http://simbad.u-strasbg.fr/simbad/sim-basic?Ident=Gaia+DR3+2217319511295868672&submit=SIMBAD+search}{BD+62  1950} & F8 & Orbital & 596.0 & 1.2e-04 & 60.7 & 0.4 & 0.5 & 0.2 & 1 & ${1.20}^{+0.07}_{-0.14}$ & $0.054$-$1.362$ \\
4598289263814843136 & \href{http://simbad.u-strasbg.fr/simbad/sim-basic?Ident=Gaia+DR3+4598289263814843136&submit=SIMBAD+search}{4598289263814843136} & N/A & Orbital & 435.8 & 1.4e-04 & 45.2 & 0.2 & 0.2 & 0.2 & 1 & ${0.71}^{+0.05}_{-0.21}$ & $0.040$-$0.726$ \\
5671384265738137984 & \href{http://simbad.u-strasbg.fr/simbad/sim-basic?Ident=Gaia+DR3+5671384265738137984&submit=SIMBAD+search}{2MASSI J0952219-192431} & M7Ve & Orbital & 278.5 & 1.5e-04 & 16.8 & 0.2 & 0.2 & 0.2 & 1 & ${0.15}^{+0.05}_{-0.08}$ & $0.046$-$0.226$ \\
2744694491118490752 & \href{http://simbad.u-strasbg.fr/simbad/sim-basic?Ident=Gaia+DR3+2744694491118490752&submit=SIMBAD+search}{BD+05  5218} & G0 & ASB1 & 248.0 & 1.7e-04 & 59.4 & 0.2 & 0.5 & 0.0 & 1 & ${0.98}^{+0.06}_{-0.06}$ & ${0.042}^{+0.005}_{-0.005}$ \\
1543883104727682176 & \href{http://simbad.u-strasbg.fr/simbad/sim-basic?Ident=Gaia+DR3+1543883104727682176&submit=SIMBAD+search}{TYC 3456-251-1} & -- & Orbital & 593.3 & 1.7e-04 & 55.8 & 0.2 & 0.2 & 0.2 & 1 & ${0.88}^{+0.05}_{-0.30}$ & $0.050$-$0.827$ \\
6604510475374443392 & \href{http://simbad.u-strasbg.fr/simbad/sim-basic?Ident=Gaia+DR3+6604510475374443392&submit=SIMBAD+search}{6604510475374443392} & N/A & Orbital & 761.5 & 1.9e-04 & 57.2 & 0.2 & 0.2 & 0.2 & 1 & ${0.88}^{+0.06}_{-0.14}$ & $0.048$-$1.025$ \\
5077267349557463680 & \href{http://simbad.u-strasbg.fr/simbad/sim-basic?Ident=Gaia+DR3+5077267349557463680&submit=SIMBAD+search}{TYC 6437-159-1} & -- & Orbital & 436.1 & 2.0e-04 & 61.3 & 0.2 & 0.2 & 0.2 & 1 & ${0.95}^{+0.06}_{-0.09}$ & $0.054$-$1.168$ \\
3087990814774129920 & \href{http://simbad.u-strasbg.fr/simbad/sim-basic?Ident=Gaia+DR3+3087990814774129920&submit=SIMBAD+search}{TYC  198-493-1} & -- & Orbital & 722.6 & 2.0e-04 & 56.1 & 0.4 & 0.5 & 0.2 & 1 & ${0.83}^{+0.05}_{-0.27}$ & $0.047$-$0.812$ \\
5488452086658908544 & \href{http://simbad.u-strasbg.fr/simbad/sim-basic?Ident=Gaia+DR3+5488452086658908544&submit=SIMBAD+search}{HD  65577} & F0IV & Orbital & 678.3 & 2.1e-04 & 85.7 & 0.2 & 0.2 & 0.2 & 1 & ${1.54}^{+0.05}_{-0.19}$ & $0.072$-$1.719$ \\
1856067303769195648 & \href{http://simbad.u-strasbg.fr/simbad/sim-basic?Ident=Gaia+DR3+1856067303769195648&submit=SIMBAD+search}{HD 340935} & G2 & Orbital & 1176.2 & 2.1e-04 & 63.3 & 0.2 & 0.5 & 0.0 & 1 & ${0.97}^{+0.06}_{-0.06}$ & ${0.058}^{+0.006}_{-0.006}$ \\
6500589824638636416 & \href{http://simbad.u-strasbg.fr/simbad/sim-basic?Ident=Gaia+DR3+6500589824638636416&submit=SIMBAD+search}{HD 219145} & F5V & Orbital & 316.0 & 2.1e-04 & 73.9 & 0.2 & 0.5 & 0.0 & 1 & ${1.21}^{+0.06}_{-0.19}$ & $0.065$-$1.342$ \\
766977465670727936 & \href{http://simbad.u-strasbg.fr/simbad/sim-basic?Ident=Gaia+DR3+766977465670727936&submit=SIMBAD+search}{HD 100360} & K0 & OTS & 572.1 & 2.3e-04 & 66.7 & 0.2 & 0.0 & 0.5 & 0 & NaN & NaN \\
4189636354105111168 & \href{http://simbad.u-strasbg.fr/simbad/sim-basic?Ident=Gaia+DR3+4189636354105111168&submit=SIMBAD+search}{TYC 5733-1115-1} & -- & ASB1 & 947.1 & 2.5e-04 & 68.5 & 0.2 & 0.0 & 0.5 & 0 & NaN & NaN \\
273199953420523648 & \href{http://simbad.u-strasbg.fr/simbad/sim-basic?Ident=Gaia+DR3+273199953420523648&submit=SIMBAD+search}{TYC 3734-75-1} & -- & Orbital & 391.5 & 2.6e-04 & 64.4 & 0.2 & 0.2 & 0.2 & 1 & ${0.88}^{+0.05}_{-0.30}$ & $0.056$-$0.854$ \\
5144739739589028992 & \href{http://simbad.u-strasbg.fr/simbad/sim-basic?Ident=Gaia+DR3+5144739739589028992&submit=SIMBAD+search}{LP  769-41} & -- & Orbital & 150.8 & 2.9e-04 & 47.2 & 0.2 & 0.2 & 0.2 & 1 & ${0.52}^{+0.05}_{-0.17}$ & $0.064$-$0.589$ \\
4224582166524567040 & \href{http://simbad.u-strasbg.fr/simbad/sim-basic?Ident=Gaia+DR3+4224582166524567040&submit=SIMBAD+search}{4224582166524567040} & N/A & Orbital & 413.7 & 2.9e-04 & 50.8 & 0.4 & 0.2 & 0.5 & 1 & ${0.58}^{+0.05}_{-0.17}$ & $0.045$-$0.657$ \\
4668056781289612928 & \href{http://simbad.u-strasbg.fr/simbad/sim-basic?Ident=Gaia+DR3+4668056781289612928&submit=SIMBAD+search}{CD-69   203} & -- & ASB1 & 429.0 & 3.1e-04 & 79.8 & 0.2 & 0.5 & 0.0 & 1 & ${1.11}^{+0.05}_{-0.11}$ & $0.070$-$1.357$ \\
6574387911224287488 & \href{http://simbad.u-strasbg.fr/simbad/sim-basic?Ident=Gaia+DR3+6574387911224287488&submit=SIMBAD+search}{L  499-75} & M3 & OTS & 240.3 & 3.3e-04 & 50.0 & 0.5 & 0.5 & 0.5 & 1 & ${0.53}^{+0.05}_{-0.18}$ & $0.045$-$0.597$ \\
1973932545092255616 & \href{http://simbad.u-strasbg.fr/simbad/sim-basic?Ident=Gaia+DR3+1973932545092255616&submit=SIMBAD+search}{TYC 3196-1933-1} & -- & Orbital & 402.0 & 3.3e-04 & 81.5 & 0.4 & 0.5 & 0.2 & 1 & ${1.11}^{+0.06}_{-0.09}$ & $0.070$-$1.393$ \\
5905528909017796480 & \href{http://simbad.u-strasbg.fr/simbad/sim-basic?Ident=Gaia+DR3+5905528909017796480&submit=SIMBAD+search}{5905528909017796480} & N/A & Orbital & 606.6 & 3.4e-04 & 72.3 & 0.4 & 0.5 & 0.2 & 1 & ${0.92}^{+0.06}_{-0.19}$ & $0.062$-$1.050$ \\
2717981134566374272 & \href{http://simbad.u-strasbg.fr/simbad/sim-basic?Ident=Gaia+DR3+2717981134566374272&submit=SIMBAD+search}{V* V335 Peg} & F5 & OTS & 787.7 & 3.5e-04 & 90.0 & 0.5 & 0.5 & 0.5 & 1 & ${1.25}^{+0.06}_{-0.14}$ & $0.079$-$1.500$ \\
2347301645624022784 & \href{http://simbad.u-strasbg.fr/simbad/sim-basic?Ident=Gaia+DR3+2347301645624022784&submit=SIMBAD+search}{CD-25   215} & -- & Orbital & 474.2 & 3.6e-04 & 78.4 & 0.2 & 0.2 & 0.2 & 1 & ${1.00}^{+0.06}_{-0.17}$ & $0.066$-$1.182$ \\
2042732453670353664 & \href{http://simbad.u-strasbg.fr/simbad/sim-basic?Ident=Gaia+DR3+2042732453670353664&submit=SIMBAD+search}{2042732453670353664} & N/A & Orbital & 497.0 & 3.6e-04 & 72.9 & 0.2 & 0.2 & 0.2 & 1 & ${0.89}^{+0.06}_{-0.14}$ & $0.062$-$1.077$ \\
6717694064902244224 & \href{http://simbad.u-strasbg.fr/simbad/sim-basic?Ident=Gaia+DR3+6717694064902244224&submit=SIMBAD+search}{TYC 7922-716-1} & -- & ASB1 & 288.7 & 3.6e-04 & 68.5 & 0.5 & 0.5 & 0.5 & 1 & ${0.81}^{+0.05}_{-0.06}$ & ${0.067}^{+0.012}_{-0.012}$ \\
4503350935885071872 & \href{http://simbad.u-strasbg.fr/simbad/sim-basic?Ident=Gaia+DR3+4503350935885071872&submit=SIMBAD+search}{TYC 1557-2079-1} & -- & Orbital & 462.6 & 3.7e-04 & 82.6 & 0.4 & 0.5 & 0.2 & 1 & ${1.07}^{+0.06}_{-0.10}$ & $0.074$-$1.362$ \\
5501729312336121856 & \href{http://simbad.u-strasbg.fr/simbad/sim-basic?Ident=Gaia+DR3+5501729312336121856&submit=SIMBAD+search}{UCAC2   9182345} & -- & ASB1 & 130.3 & 3.8e-04 & 62.2 & 0.5 & 0.5 & 0.5 & 1 & ${0.68}^{+0.05}_{-0.05}$ & ${0.061}^{+0.010}_{-0.010}$ \\
2214819118774573440 & \href{http://simbad.u-strasbg.fr/simbad/sim-basic?Ident=Gaia+DR3+2214819118774573440&submit=SIMBAD+search}{BD+68  1377} & G0 & ASB1 & 310.9 & 4.2e-04 & 81.3 & 0.2 & 0.5 & 0.0 & 1 & ${0.97}^{+0.06}_{-0.06}$ & ${0.081}^{+0.009}_{-0.010}$ \\
4474368187335444352 & \href{http://simbad.u-strasbg.fr/simbad/sim-basic?Ident=Gaia+DR3+4474368187335444352&submit=SIMBAD+search}{TYC  438-609-1} & -- & Orbital & 508.1 & 4.3e-04 & 79.4 & 0.4 & 0.5 & 0.2 & 1 & ${0.93}^{+0.05}_{-0.14}$ & $0.068$-$1.143$ \\
1378974777381950080 & \href{http://simbad.u-strasbg.fr/simbad/sim-basic?Ident=Gaia+DR3+1378974777381950080&submit=SIMBAD+search}{TYC 3061-368-1} & -- & Orbital & 309.3 & 4.5e-04 & 77.2 & 0.2 & 0.5 & 0.0 & 1 & ${0.87}^{+0.05}_{-0.30}$ & $0.069$-$0.837$ \\
6827680034091744896 & \href{http://simbad.u-strasbg.fr/simbad/sim-basic?Ident=Gaia+DR3+6827680034091744896&submit=SIMBAD+search}{HD 203771} & K1V & OTS & 1105.4 & 4.5e-04 & 73.0 & 0.2 & 0.2 & 0.2 & 1 & ${0.80}^{+0.05}_{-0.19}$ & $0.065$-$0.920$ \\
4984264505688458112 & \href{http://simbad.u-strasbg.fr/simbad/sim-basic?Ident=Gaia+DR3+4984264505688458112&submit=SIMBAD+search}{UCAC4 245-001421} & -- & ASB1 & 196.7 & 4.6e-04 & 80.0 & 0.5 & 0.5 & 0.5 & 1 & ${0.91}^{+0.06}_{-0.06}$ & ${0.075}^{+0.009}_{-0.010}$ \\
6228661976111571072 & \href{http://simbad.u-strasbg.fr/simbad/sim-basic?Ident=Gaia+DR3+6228661976111571072&submit=SIMBAD+search}{HD 131197} & K2III/IV & Orbital & 421.0 & 4.8e-04 & 79.0 & 0.4 & 0.2 & 0.5 & 1 & ${0.87}^{+0.05}_{-0.15}$ & $0.070$-$1.063$ \\
5538678297515561088 & \href{http://simbad.u-strasbg.fr/simbad/sim-basic?Ident=Gaia+DR3+5538678297515561088&submit=SIMBAD+search}{CD-38  3568} & -- & ASB1 & 267.6 & 5.0e-04 & 87.4 & 0.5 & 0.5 & 0.5 & 1 & ${1.00}^{+0.06}_{-0.07}$ & ${0.088}^{+0.013}_{-0.014}$ \\
4390684426062198528 & \href{http://simbad.u-strasbg.fr/simbad/sim-basic?Ident=Gaia+DR3+4390684426062198528&submit=SIMBAD+search}{TYC  413-602-1} & -- & ASB1 & 240.9 & 5.1e-04 & 86.5 & 0.2 & 0.5 & 0.0 & 1 & ${0.96}^{+0.06}_{-0.07}$ & $0.073$-$1.280$ \\
1693282786205650560 & \href{http://simbad.u-strasbg.fr/simbad/sim-basic?Ident=Gaia+DR3+1693282786205650560&submit=SIMBAD+search}{2MASS J12114960+7752445} & -- & Orbital & 651.2 & 5.2e-04 & 54.3 & 0.2 & 0.2 & 0.2 & 1 & ${0.47}^{+0.05}_{-0.17}$ & $0.048$-$0.548$ \\
6186817449775754496 & \href{http://simbad.u-strasbg.fr/simbad/sim-basic?Ident=Gaia+DR3+6186817449775754496&submit=SIMBAD+search}{PM J13128-2603} & -- & ASB1 & 87.0 & 5.2e-04 & 64.2 & 0.2 & 0.2 & 0.2 & 1 & ${0.60}^{+0.05}_{-0.05}$ & ${0.075}^{+0.013}_{-0.013}$ \\
6345830951689788672 & \href{http://simbad.u-strasbg.fr/simbad/sim-basic?Ident=Gaia+DR3+6345830951689788672&submit=SIMBAD+search}{CPD-85   499} & G5 & ASB1 & 485.0 & 5.7e-04 & 92.8 & 0.2 & 0.5 & 0.0 & 1 & ${1.01}^{+0.06}_{-0.07}$ & ${0.080}^{+0.012}_{-0.012}$ \\
5829002521785164160 & \href{http://simbad.u-strasbg.fr/simbad/sim-basic?Ident=Gaia+DR3+5829002521785164160&submit=SIMBAD+search}{TYC 9040-2811-1} & -- & ASB1 & 508.3 & 6.4e-04 & 107.3 & 0.2 & 0.5 & 0.0 & 1 & ${1.19}^{+0.06}_{-0.06}$ & ${0.087}^{+0.014}_{-0.015}$ \\
4188152322648398592 & \href{http://simbad.u-strasbg.fr/simbad/sim-basic?Ident=Gaia+DR3+4188152322648398592&submit=SIMBAD+search}{2MASS J19262435-1045398} & -- & Orbital & 175.8 & 7.0e-04 & 49.8 & 0.2 & 0.2 & 0.2 & 1 & ${0.34}^{+0.05}_{-0.16}$ & $0.043$-$0.422$ \\
548478735430284416 & \href{http://simbad.u-strasbg.fr/simbad/sim-basic?Ident=Gaia+DR3+548478735430284416&submit=SIMBAD+search}{BD+74   118} & K5V & OTS & 48.1 & 7.6e-04 & 86.9 & 0.4 & 0.5 & 0.2 & 1 & ${0.78}^{+0.05}_{-0.21}$ & $0.078$-$0.920$ \\
5344309621263564160 & \href{http://simbad.u-strasbg.fr/simbad/sim-basic?Ident=Gaia+DR3+5344309621263564160&submit=SIMBAD+search}{TYC 8631-835-1} & -- & ASB1 & 789.9 & 7.7e-04 & 97.0 & 0.2 & 0.5 & 0.0 & 1 & ${0.92}^{+0.05}_{-0.06}$ & ${0.078}^{+0.009}_{-0.009}$ \\
6012944872855490944 & \href{http://simbad.u-strasbg.fr/simbad/sim-basic?Ident=Gaia+DR3+6012944872855490944&submit=SIMBAD+search}{HD 139096} & G3V & ASB1 & 129.5 & 8.3e-04 & 104.6 & 0.2 & 0.2 & 0.2 & 1 & ${1.00}^{+0.06}_{-0.06}$ & ${0.079}^{+0.009}_{-0.009}$ \\
4441168914767151360 & \href{http://simbad.u-strasbg.fr/simbad/sim-basic?Ident=Gaia+DR3+4441168914767151360&submit=SIMBAD+search}{HD 153402} & K0 & OTS & 104.2 & 8.4e-04 & 96.1 & 0.2 & 0.0 & 0.5 & 0 & ${0.87}^{+0.05}_{-0.19}$ & $0.088$-$1.071$ \\
3063883403860016512 & \href{http://simbad.u-strasbg.fr/simbad/sim-basic?Ident=Gaia+DR3+3063883403860016512&submit=SIMBAD+search}{TYC 4858-1465-1} & -- & ASB1 & 328.7 & 9.4e-04 & 102.4 & 0.2 & 0.0 & 0.5 & 0 & ${0.90}^{+0.05}_{-0.05}$ & ${0.101}^{+0.015}_{-0.016}$ \\
4725142772167859840 & \href{http://simbad.u-strasbg.fr/simbad/sim-basic?Ident=Gaia+DR3+4725142772167859840&submit=SIMBAD+search}{L  127-40} & -- & ASB1 & 118.8 & 1.1e-03 & 96.2 & 0.2 & 0.2 & 0.2 & 1 & ${0.74}^{+0.06}_{-0.05}$ & ${0.084}^{+0.013}_{-0.012}$ \\
3067929881523376384 & \href{http://simbad.u-strasbg.fr/simbad/sim-basic?Ident=Gaia+DR3+3067929881523376384&submit=SIMBAD+search}{TYC 4854-2258-1} & -- & ASB1 & 262.7 & 1.1e-03 & 117.0 & 0.2 & 0.0 & 0.5 & 0 & ${0.99}^{+0.06}_{-0.06}$ & ${0.182}^{+0.040}_{-0.038}$ \\
1003812373374277248 & \href{http://simbad.u-strasbg.fr/simbad/sim-basic?Ident=Gaia+DR3+1003812373374277248&submit=SIMBAD+search}{HD  49039} & G5 & OTS & 19.9 & 1.5e-03 & 117.4 & 0.2 & 0.0 & 0.5 & 0 & ${0.87}^{+0.06}_{-0.21}$ & $0.100$-$1.122$ \\
4955501899980809088 & \href{http://simbad.u-strasbg.fr/simbad/sim-basic?Ident=Gaia+DR3+4955501899980809088&submit=SIMBAD+search}{TYC 7548-325-1} & -- & ASB1 & 337.3 & 1.8e-03 & 125.2 & 0.2 & 0.2 & 0.2 & 1 & ${0.86}^{+0.06}_{-0.06}$ & ${0.082}^{+0.013}_{-0.013}$ \\
5247111556405603840 & \href{http://simbad.u-strasbg.fr/simbad/sim-basic?Ident=Gaia+DR3+5247111556405603840&submit=SIMBAD+search}{TYC 8952-212-1} & -- & ASB1 & 402.4 & 2.2e-03 & 126.2 & 0.2 & 0.5 & 0.0 & 1 & ${0.76}^{+0.06}_{-0.05}$ & ${0.098}^{+0.020}_{-0.019}$ \\
5612492292661869824 & \href{http://simbad.u-strasbg.fr/simbad/sim-basic?Ident=Gaia+DR3+5612492292661869824&submit=SIMBAD+search}{HD  60666} & K1III & OTS & 29.1 & 2.3e-03 & 150.3 & 0.2 & 0.0 & 0.5 & 0 & NaN & NaN \\
2743198605549406464 & \href{http://simbad.u-strasbg.fr/simbad/sim-basic?Ident=Gaia+DR3+2743198605549406464&submit=SIMBAD+search}{HD 223238} & G5V & OTS & 61.3 & 2.5e-03 & 157.9 & 0.2 & 0.0 & 0.5 & 0 & ${1.01}^{+0.06}_{-0.10}$ & $0.133$-$1.364$ \\
6911090115750857344 & \href{http://simbad.u-strasbg.fr/simbad/sim-basic?Ident=Gaia+DR3+6911090115750857344&submit=SIMBAD+search}{SCR J2108-0600} & -- & ASB1 & 144.1 & 2.9e-03 & 144.0 & 0.2 & 0.5 & 0.0 & 1 & ${0.81}^{+0.05}_{-0.05}$ & ${0.094}^{+0.015}_{-0.015}$ \\
798068905726303232 & \href{http://simbad.u-strasbg.fr/simbad/sim-basic?Ident=Gaia+DR3+798068905726303232&submit=SIMBAD+search}{*  11 LMi} & G8Va & OTS & 27.1 & 3.0e-03 & 156.2 & 0.2 & 0.0 & 0.5 & 0 & ${0.91}^{+0.05}_{-0.11}$ & $0.137$-$1.182$ \\
6795834500861297408 & \href{http://simbad.u-strasbg.fr/simbad/sim-basic?Ident=Gaia+DR3+6795834500861297408&submit=SIMBAD+search}{SIPS J2049-2800} & sdM6.5? & Orbital & 641.1 & 3.1e-03 & 52.0 & 0.2 & 0.2 & 0.2 & 1 & ${0.15}^{+0.05}_{-0.07}$ & $0.065$-$0.348$ \\
3931991608291720192 & \href{http://simbad.u-strasbg.fr/simbad/sim-basic?Ident=Gaia+DR3+3931991608291720192&submit=SIMBAD+search}{TYC  880-843-1} & -- & ASB1 & 257.6 & 3.4e-03 & 184.5 & 0.2 & 0.5 & 0.0 & 1 & ${1.10}^{+0.06}_{-0.06}$ & ${0.093}^{+0.016}_{-0.015}$ \\
845583819682861184 & \href{http://simbad.u-strasbg.fr/simbad/sim-basic?Ident=Gaia+DR3+845583819682861184&submit=SIMBAD+search}{LSPM J1131+5627} & -- & Orbital & 392.5 & 1.3e-02 & 84.1 & 0.2 & 0.2 & 0.2 & 1 & ${0.12}^{+0.05}_{-0.05}$ & ${0.080}^{+0.019}_{-0.018}$ \\
5207515806222756480 & \href{http://simbad.u-strasbg.fr/simbad/sim-basic?Ident=Gaia+DR3+5207515806222756480&submit=SIMBAD+search}{CPD-81   193} & -- & ASB1 & 71.8 & 1.5e-02 & 292.0 & 0.2 & 0.5 & 0.0 & 1 & ${0.92}^{+0.05}_{-0.05}$ & ${0.080}^{+0.017}_{-0.016}$ \\

\hline
\end{tabular}
\end{table*}

\begin{figure} 
\centering
\includegraphics[width=\linewidth]{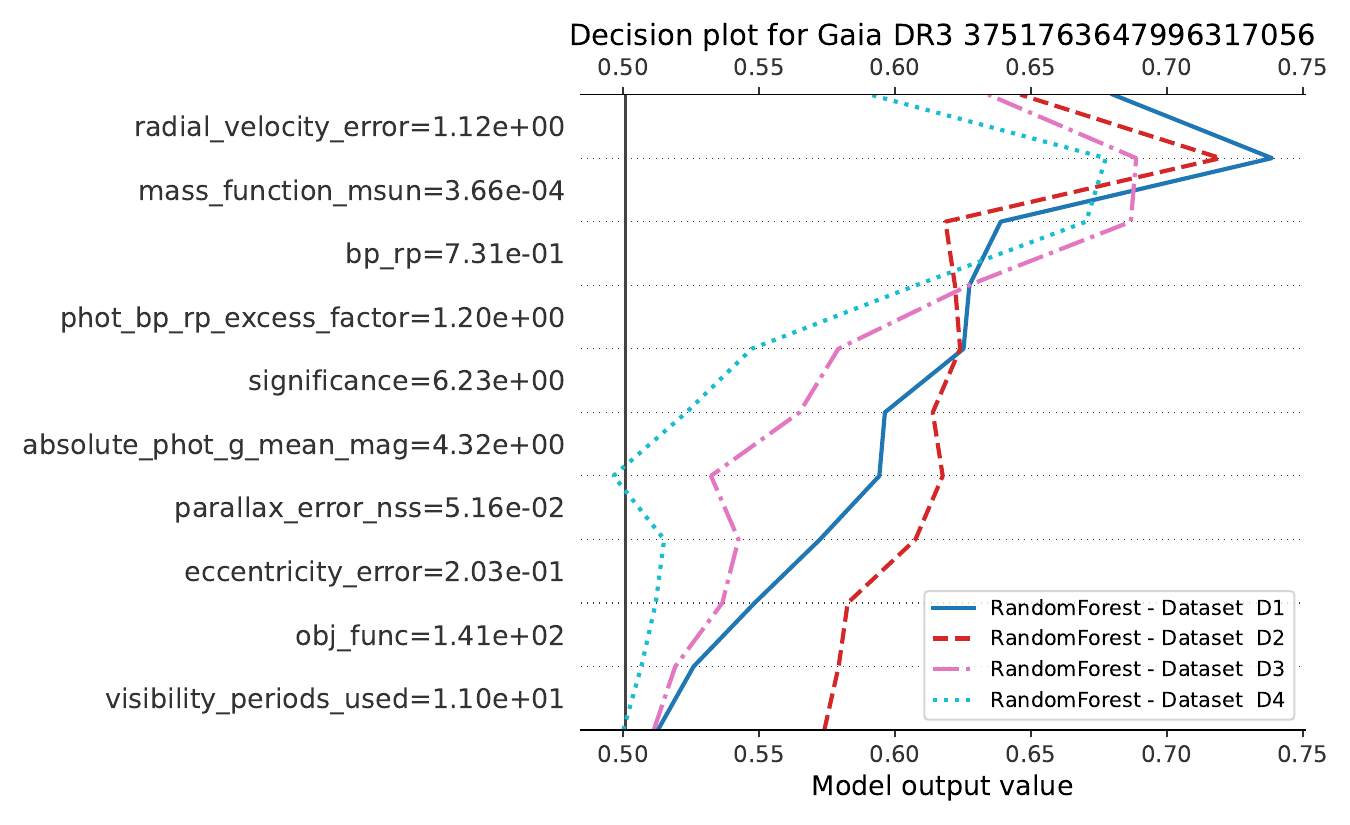}
\caption{Decision plot showing feature contribution to model output for the random forest models for the brown dwarf in HD 89707. Radial-velocity error and mass\_function\_msun were the most decisive features where the most important features. \label{fig:feat_bd2_rf}}
\end{figure}

\begin{figure} 
\centering
\includegraphics[width=\linewidth]{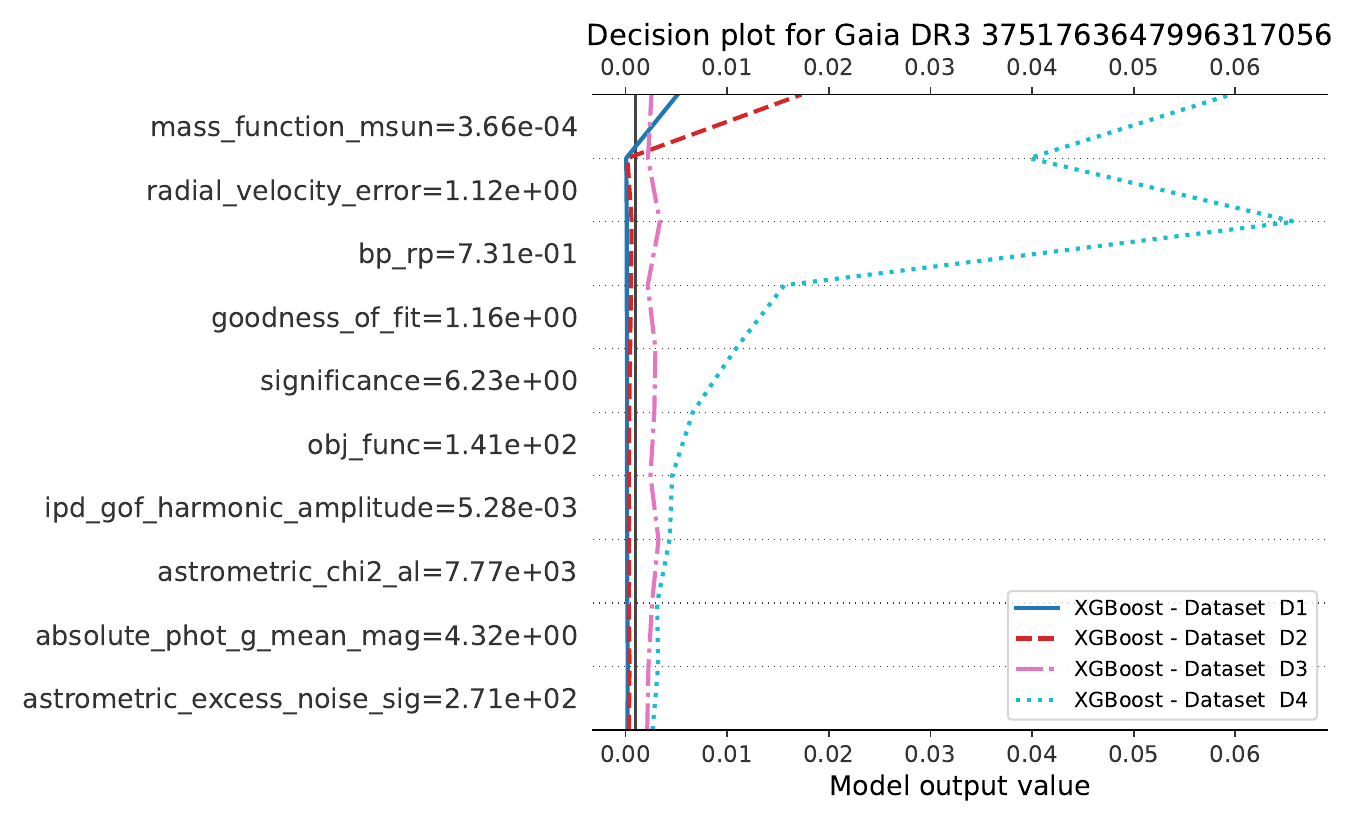}
\caption{Decision plot showing feature contribution to model output for the XGBoost models for the brown dwarf in HD 89707. Radial-velocity error and mass\_function\_msun were the most decisive features where the most important features.  \label{fig:feat_bd2_xgb}}
\end{figure}

% If you want to present additional material which would interrupt the flow of the main paper,
% it can be placed in an Appendix which appears after the list of references.

%%%%%%%%%%%%%%%%%%%%%%%%%%%%%%%%%%%%%%%%%%%%%%%%%%

% Don't change these lines
\bsp	% typesetting comment
\label{lastpage}
\end{document}